\documentclass[pdflatex,sn-mathphys-num]{sn-jnl}

\usepackage{graphicx}%
\usepackage{multirow}%
\usepackage{amsmath,amssymb,amsfonts}%
\usepackage{amsthm}%
\usepackage{mathrsfs}%
\usepackage[title]{appendix}%
\usepackage{xcolor}%
\usepackage{textcomp}%
\usepackage{manyfoot}%
\usepackage{booktabs}%
\usepackage{algorithm}%
\usepackage{algorithmicx}%
\usepackage{algpseudocode}%
\usepackage{listings}%
\usepackage{svg}%
\usepackage{subcaption}%
\usepackage{pgfplots}%
\usepackage{siunitx}%
\usepackage{makecell}

\theoremstyle{thmstyleone}%
\theoremstyle{thmstyletwo}%

\theoremstyle{thmstylethree}%

\definecolor{backcolor}{rgb}{0.95,0.95,0.92}
\lstdefinestyle{mystyle}{
    backgroundcolor=\color{backcolor},
    numberstyle=\tiny,
    basicstyle=\fontsize{6}{9}\selectfont\ttfamily,
    breakatwhitespace=false,
    breaklines=true,
    captionpos=b,
    keepspaces=true,
    numbers=left,
    numbersep=5pt,
    showspaces=false,
    showstringspaces=false,
    showtabs=false,
    tabsize=2
}
\usepackage[]{fixme}
\FXRegisterAuthor{vs}{evs}{VS$\!\!$} 
\FXRegisterAuthor{td}{TODO}{\color{red}\tiny{\bf todo}}
\FXRegisterAuthor{is}{eis}{IS}

\fxsetup{targetlayout=color}
\fxsetface{margin}{\tiny}
\fxsetup{envface=\tiny\it}  
\fxsetup{inlineface=\tiny\bfseries}

\makeatletter
\newcommand{\onelettername}[1]{#1\aftergroup\@gobble}
\makeatother
\usepackage[normalem]{ulem}

\begin{document}

\title[Article Title]{Ruleless Digital Twins: Toward Declarative Decision-Making Through Standardized Frameworks and Technologies}


\author*{\fnm{Ivan} \sur{Spajić}}\email{ispa@hvl.no}

\author{\fnm{Volker} \sur{Stolz}}\email{vsto@hvl.no}

\affil{\orgdiv{Department of Computer Science, Electrical Engineering, and Mathematical Sciences}, \orgname{Western Norway University of Applied Sciences}, \city{Bergen}, \country{Norway}}




\abstract{Digital twins (DTs) can be thought of as digital counterparts of physical objects, or more generally, twinning targets (TTs).
To enact changes on and optimize for properties of their TTs, DTs use an element of decision-making.
Traditionally, many domains, such as home automation, utilize rule-based decision-making models to imperatively define DT actions based on desired TT conditions.
With evolving user specifications, rule-based models have become increasingly complex to develop and maintain, particularly in scenarios involving optimizations under dynamically changing systems, such as those affected by weather or dynamic energy pricing.
We present an alternative in the form of \textit{ruleless digital twins} (RDTs) that automatically produce optimal decisions with respect to purely declarative user specifications, similarly to the well-established \textit{ruleless} approach of model predictive control.
They do so through a combination of a semantic knowledge base, logical inference, simulation models, and an autonomic computing architecture, each exclusively based on a widely-used standard or technology.
We evaluate our proof of concept through an incubator case study and experiments against a rule-based bang-bang controller in the context of a virtual office room environment.
Results show core that RDT functionality performs better than the control in terms of maintaining desired room temperatures and minimizing energy costs under dynamically changing spot pricing.
Additionally, due to combinatorial decision tree construction, RDTs require considerable computational power for longer prediction horizons, although their applicability to specific problem domains (and use of shorter horizons) ultimately falls on their users.}




\maketitle

\section{Introduction}
\label{introduction}
Digital twins (DTs) are commonly defined as ``digital counterparts of physical objects'' \cite{kritzinger_et_al_2018}.
Although often varied, definitions involving DTs usually include the three main elements of a cyber-physical system: (\textit{i}) an object or process, traditionally known as the physical twin (PT) and more recently as the twinning target (TT) \cite{lehner_et_al_2025}, (\textit{ii}) a digital element representing it, known as the DT, and (\textit{iii}) some form of data transfer between the two \cite{digital_twin_consortium, kritzinger_et_al_2018, larsen_et_al_2024}.
Crucially, Larsen \textit{et al.} \cite{larsen_et_al_2024} also specify that a DT should provide \textit{value} to its TT\footnote{Larsen et al. use the term ``PT''.}, thus serving a tangible, user-oriented purpose.
Value in this context can be considered a general improvement or, in more technical terms, a kind of optimization with respect to user specifications.

To achieve such optimizations, DTs must transmit data to their TTs to enact certain real-world changes.
Doing so therefore involves an element of decision-making, whether it is done by a human-in-the-loop supported by the DT or the DT itself \cite{coelho_et_al_2021, galvez_del_postigo_gallego_et_al_2025, walton_et_al_2024}.
For example, DTs have been used as decision support systems (DSSs) facilitating decision-making in various domains, such as planning \cite{magnanini_et_al_2021}, production and distribution \cite{coelho_et_al_2021}, and warehouse optimization \cite{korth_et_al_2018}.
For full, unassisted decision-making, DTs have utilized rule-based models commonly consisting of explicit rules \cite{bolender_et_al_2021, dalibor_et_al_2022, michael_et_al_2024, yigal_et_al_2024} or finite-state automata \cite{compagnucci_et_al_2023}.
The rules in such models typically follow the \texttt{TRIGGER-CONDITION-ACTION} \cite{yigal_et_al_2024} rule paradigm and are commonly used in domains like home automation (HA) \cite{manca_et_al_2019, newcomb_et_al_2017, openhab, soares_et_al_2021}.
They allow users to intuitively and explicably specify their trigger conditions and consequent actions, such as 
\begin{center}
    \texttt{IF roomTemperature < 18C THEN heater.state = ON}.
\end{center}
Therefore, depending on the size of the modeled TT, a respective DT with decision-making functionality in the HA domain might use models containing dozens (if not hundreds) of such rules.

Somewhat obviously, as the number of rules in a decision-making model grows, so does the complexity of managing it.
With every rule added, it becomes easier to run into different kinds of conflicts \cite{manca_et_al_2019, newcomb_et_al_2017}.
For example, users may define a pair of rules where the actions of one trigger the conditions of the other, and vice-versa, subsequently leading to unintended rule-firing loops.
Although these and other conflicts may be identified and remedied, doing so typically results in even greater numbers of ever more specific rules \cite{manca_et_al_2019}.
In addition, more complex user specifications increase model complexity even further.
Despite users in the HA domain generally using relatively simple rules \cite{soares_et_al_2021, ur_et_al_2014}, such as \texttt{"when I enter the garage, turn the light on"}, more complex needs may arise \cite{soares_et_al_scenarios_2021}.
For instance, homeowners may have multiple heating sources, such as gas-powered heaters, electrical floor heating, and heat pumps, and may wish to maintain warmth with any combination of them while simultaneously minimizing their energy bills.
Thus, it becomes apparent that specifying imperative rules for various optimizations, while also avoiding conflicts, easily turns into a non-trivial task \cite{bolender_et_al_2021}.
Additionally, during the development of the rules (at design time), it may generally be difficult to predict problems that could occur during their execution (at runtime) \cite{manca_et_al_2019}.

\subsection{Problem Domain}
\label{problem_domain}
It is important to clarify that although we emphasize some problems of imperative decision-making rules, not all such rules ought to be phased out.
Rules like \texttt{"when I enter the garage, turn the light on"} represent relatively simple problems with direct solutions that offer little room for added value \cite{larsen_et_al_2024}.
Because cases like this leave essentially no hypothetical scenarios to explore or possible optimizations to pursue, the use of DTs to this end is rather unwarranted.\footnote{Although DTs could still be used for smart home visualizations or state-tracking, the decision models fulfilling these purposes would be vastly different from those actuating smart home devices.}
Therefore, for many simple scenarios, imperative rules still remain an intuitive and explainable decision-making solution \cite{bolender_et_al_2021, manca_et_al_2019, newcomb_et_al_2017}.

We focus specifically on providing an alternative to rule-based decision-making models, whose rules attempt to handle many (if not all) possible TT states, including both controlled (actuator inputs) and observed (e.g., measured temperature) variables, particularly through optimized action choices.
To precisely illustrate the potential complexity and issues in managing such models, we present a concrete example from the HA domain.

\paragraph{Rule-Based Problem Example}
Given a simple smart home with just two binary-state actuators, an electric heater and floor heating, a user can easily construct a rudimentary, rule-based decision-making model for regulating indoor temperature. For instance, they may start with the following `bang-bang' controller that executes its rules in periodic cycles (e.g., 15 minutes):

\begin{center}
    \begin{tabular}{l | l}
        \texttt{RULE 1} & \texttt{IF indoor.temperature < 18C THEN heater.state = ON} \\
        \texttt{RULE 2} & \texttt{IF indoor.temperature > 22C THEN heater.state = OFF}
    \end{tabular}
\end{center}

Assuming the user cares about simultaneously minimizing energy consumption, they may expand the model further:

\begin{center}
    \begin{tabular}{l | l}
        \texttt{RULE 1} & \texttt{IF indoor.temperature < 16C THEN heater.state = ON} \\
        \texttt{RULE 2} & \texttt{\makecell[l]{IF indoor.temperature >= 16C AND < 18C THEN \\ \hspace{1cm} floorHeating.state = ON}} \\
        \texttt{RULE 3} & \texttt{\makecell[l]{IF indoor.temperature > 22C THEN heater.state = OFF AND \\ \hspace{1cm} floorHeating.state = OFF}} \\
        \texttt{RULE 4} & \texttt{\makecell[l]{IF energy.consumption > 10kWh THEN heater.state = OFF AND \\ \hspace{1cm} floorHeating.state = OFF}}
    \end{tabular}
\end{center}

This, however, immediately leads to two issues: firstly, there is a conflict between rules 1 and 2 and rule 4.
For example, in case the indoor temperature is less than 16$^{\circ}$C while the total energy consumption has already surpassed 10 kWh, the decision-making model demands the heater to be both on and off.
Secondly, even if the user knows about the actuators' energy consumptions and heating abilities, manually calibrating the rules (i.e., what actuator to use at what temperature) to minimize energy consumption, while also achieving the desired temperature, is certainly not trivial.
Not only may this require a trial-and-error approach to finding the optimal rule configuration, but the configuration itself may change based on dynamic factors influencing the smart home (e.g., the weather).
Naturally, expanding this simple decision-making model to cover real-world examples of personal homes or office buildings only exacerbates the highlighted problems.

\subsection{Model Predictive Control}
\label{model_predictive_control}
\begin{figure}[tbp]
    \centering
    \includegraphics[width=1\linewidth]{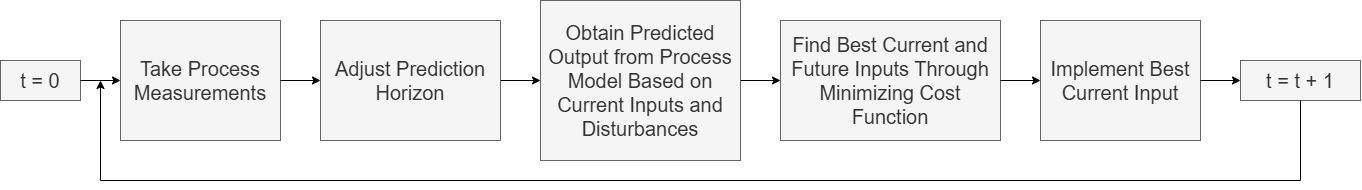}
    \caption{The iterative process of model predictive control (MPC) \cite{nikolaou_model_2001}.}
    \label{fig:mpc_loop}
\end{figure}

In the 1970s, a formal method of process control emerged in the chemical engineering industry.
Model predictive control (MPC) \cite{camacho_model_2007, nikolaou_model_2001} uses mathematical optimization techniques to calculate long-term optimal process inputs (e.g., actuator states) with respect to user specifications.
It relies on process models to predict observable variable outputs (e.g., measured temperature values) over a finite future horizon stemming from a point in time, \(t\).
By predicting outputs, it is able to generate a sequence of optimal inputs for a portion of that horizon before immediately applying those meant for \(t + 1\).
At \(t + 1\), measurements of outputs resulting from the previously applied inputs (as well as any dynamic disturbances) are taken, the prediction horizon is shifted forward, and optimal inputs are calculated once again.
MPC's full iterative process is illustrated in Figure~\ref{fig:mpc_loop}.

To find optimal process inputs, MPC relies on minimizing a cost function, traditionally expressed in the following form\footnote{Different MPC approaches and algorithms may use augmented versions of the expression. For more information on different cost functions (and process disturbance modeling), the reader is directed to Nikolaou (2001) \cite{nikolaou_model_2001} and Camacho and Bordons (2007) \cite{camacho_model_2007}.} \cite{nikolaou_model_2001}:
\begin{equation}
    J = min_{u[k|k],...,u[k + p - 1|k]} \ \left( \sum^{p}_{i = 1}w_{i}(y[k + i|k] - y_{sp})^{2} + \sum^{m}_{i = 1}r_{i}\nabla u[k + i - 1|k]^{2} \right)
\end{equation}
where \(u\) represents the process input, \(y\) the process output, \(y_{sp}\) the desired output, \(\nabla\) the backward difference operator, \(k\) the points in time, \(m\) and \(p\) the lengths of the predicted process input and output horizons, respectively, and \(w\) and \(r\) the tuning weights for more specific control strategies.
The \(z[k + i|k]\) notation depicts what the value of \(z\) will be at time \(k + i\) as predicted at time \(k\).
Finding optimal input values thus reduces to a quadratic programming problem subject to certain input and output constraints.
Such constraints may, for example, represent ranges of possible actuator states (including how drastically these may change) as well as tolerable process output values \cite{camacho_model_2007, nikolaou_model_2001}:
\begin{equation}
    u_{max} \geq u[k + i - 1|k] \geq u_{min} \ \mid \ i = 1, ..., m
\end{equation}
\begin{equation}
    \nabla u_{max} \geq \nabla u[k + i - 1|k] \geq -\nabla u_{max} \ \mid \ i = 1, ..., m
\end{equation}
\begin{equation}
    y_{max} \geq y[k + i|k] \geq y_{min} \ \mid \ i = 1, ..., p
\end{equation}

Crucially, MPC only requires users (engineers) to provide their constraints and desired objectives \textit{declaratively}.
As a result, it exercises optimized control with respect to user specifications \textit{without} the need for imperative, decision-making rule models.

\paragraph{Mixed-Integer Model Predictive Control}
\label{mixed-integer_model_predictive_control}

\begin{figure}[tbp]
    \centering
    \includegraphics[width=1\linewidth]{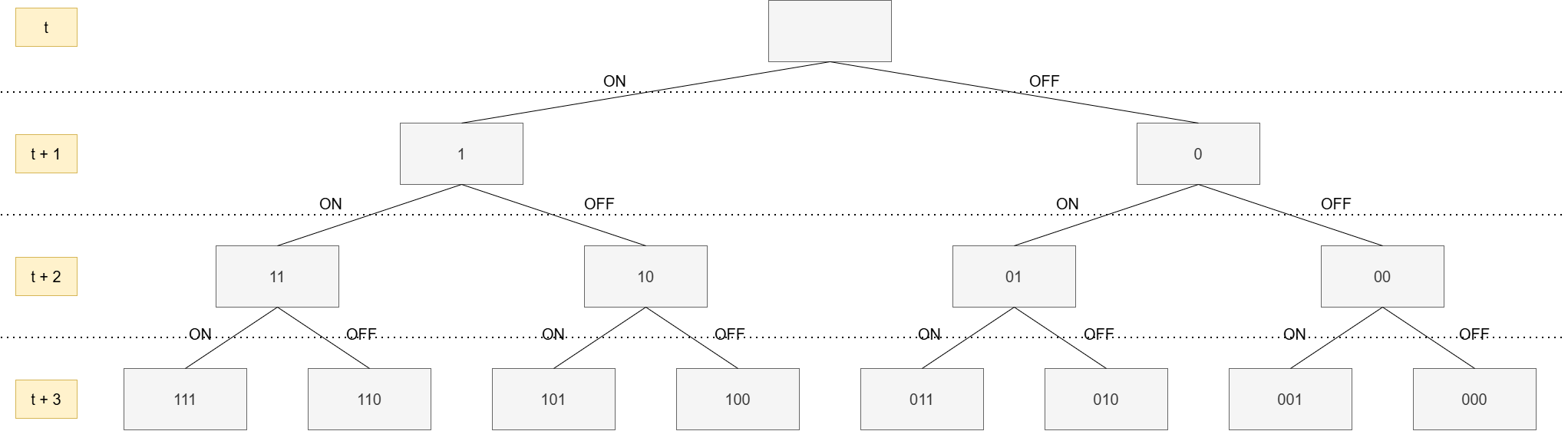}
    \caption{A tree of possible control decisions for a binary-state actuator over a 4-point horizon \cite{camacho_model_2007}.}
    \label{fig:mimpc_tree}
\end{figure}

Traditionally, problems tackled by MPC involve process inputs with continuous ranges of real (floating-point) values.
However, certain domains, such as heating, ventilation, and air conditioning (HVAC), may also contain actuators with discrete (integer-valued) states, which places additional constraints on optimizations.
Over the duration of a given prediction horizon, different combinations of discrete actuator states form a tree of possible control decisions, as illustrated by the example in Figure~\ref{fig:mimpc_tree}.
Due to the discrete nature of possible input combinations, obtaining the optimal one through so-called mixed-integer programming is mathematically more difficult.
As the tree scales exponentially with the number of actuator states, performing a brute-force search quickly becomes computationally infeasible.
To perform its optimizations, mixed-integer MPC thus relies on heuristics based on various branch-and-bound algorithms \cite{camacho_model_2007, mcallister_advances_2022, hariprasad_efficient_2017}.

\subsection{Contributions}
\label{contributions}

\begin{figure}[tbp]
    \centering
    \includegraphics[width=0.75\linewidth]{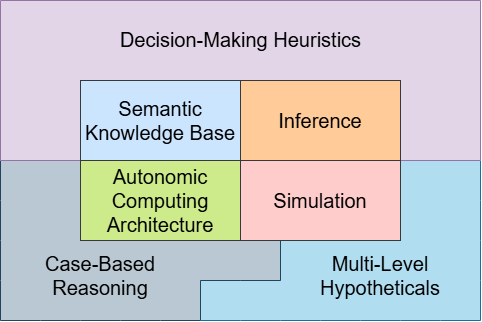} 
    \caption{A diagram of concepts that make up a ruleless digital twin (RDT). Specific secondary elements build on specific parts of the core.}
    \label{fig:contributions}
\end{figure}

In this work, we present and evaluate \textit{ruleless digital twins} (RDTs) demonstrating autonomous TT control with respect to user specifications without the use of imperative decision-making models.
We update and extend our earlier idea and prototype \cite{spajic_and_stolz_2026} with more comprehensive inference logic, dynamic decision tree construction, a functional decision ranking-and-selection algorithm \cite{walton_et_al_2024, hong_et_al_2021}, and additional, secondary features optimizing decision-making and performance (Figure~\ref{fig:contributions}).
All of the methods and techniques introduced in this section are discussed in detail in Section~\ref{method}.

\paragraph{Primary Contribution}
Our primary contribution is the demonstration of how widely-adopted standards and technologies, consisting of a semantic knowledge base, logical inference, simulations, and an autonomic computing architecture, combine to form a DT-based controller capable of enacting optimized actions, akin to those of mixed-integer MPC, without using imperative, decision-making rules.

As a result, by only requiring declarative specifications and no imperative rules, \textit{ruleless} DTs ultimately relieve end users of complex decision model development, albeit at initially higher computational costs (Sections~\ref{semantic_knowledge_base}~-~\ref{mape-k_loop}).

\paragraph{Secondary Contributions}
To enhance our primary contribution, we utilize several additional techniques non-essential to core RDT functionality:
\begin{itemize}
    \item Instead of obtaining optimal decisions through full decision tree (Figure~\ref{fig:mimpc_tree}) searches, RDTs may instead employ reactive decision-making heuristics to improve performance and reduce cycle times by addressing only the immediate \textit{needs} of their TTs (Sections~\ref{semantic_knowledge_base},~\ref{logical_inference},~and~\ref{mape-k_loop}).
    \item Although RDTs utilize simulations to explore possible scenarios, they can do so in a nested manner.
        Such multi-level hypotheticals (of hypotheticals) enable them to make potentially suboptimal short-term choices which may yield greater long-term benefits (Section~\ref{multi_level_hypotheticals}) \cite{nikolaou_model_2001}.
    \item After running for some time, it is possible that an RDT observes previously-encountered TT conditions (property values) once again.
    If user specifications remain unchanged, the RDT should not have to (re-)simulate to (re-)produce appropriate decisions already taken.
    Hence, we make use of case-based reasoning \cite{aamodt_and_plaza_1994} to trade computational cost against storage by recording and reusing TT conditions and their respective, successfully mitigating decisions (Section~\ref{case_based_reasoning}).
\end{itemize}

The rest of the paper is structured as follows: in Section~\ref{method}, we outline the different components of an RDT and how different parts of it work.
We also explain the inner workings of our secondary contributions and the improvements they provide.
This is followed by Section~\ref{validation}, where we use two scenarios to validate our proof-of-concept RDT experimentally and discuss the results.
Section~\ref{discussion} discusses the pros and cons of RDTs, and Section~\ref{related_work} outlines the related work.
Finally, in Sections~\ref{future_work} and~\ref{conclusion}, we present possible paths for future work and conclude, respectively.
  

\section{The \textit{Ruleless} Method}
\label{method}
\begin{figure}[tbp]
    \centering
    \includegraphics[width=1\linewidth]{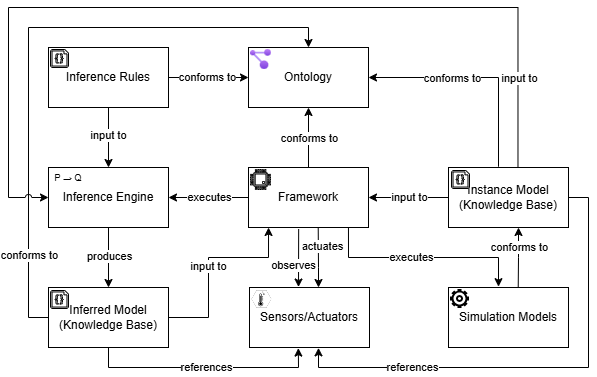}
    \caption{An architectural component diagram of our proof-of-concept RDT.}
    \label{fig:framework_architecture}
\end{figure}

\begin{figure}[tbp]
    \centering
    \includegraphics[width=1\linewidth]{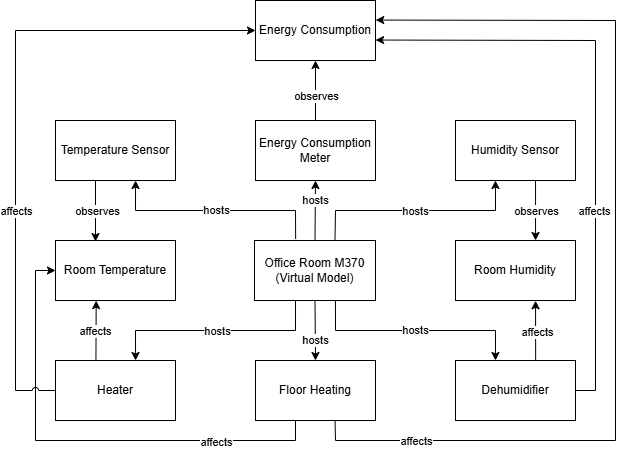}
    \caption{A component diagram of virtual office room M370.}
    \label{fig:office_room_m370}
\end{figure}

The following section details each of the components comprising the core functionality of an RDT as well as the non-essential techniques improving its effectiveness.
The architectural components of our proof-of-concept implementation\footnote{All source files and validation results are available at \url{https://github.com/ivanspajic/ruleless-digital-twins} and will be archived on e.g. Zenodo for the published version.} are illustrated in Figure~\ref{fig:framework_architecture}.
For added clarity, we will use a virtual model of our office room M370 as a running example of a TT, whose most relevant IoT components are shown in Figure~\ref{fig:office_room_m370}.

\subsection{Semantic Knowledge Base}
\label{semantic_knowledge_base}
The two most powerful aspects of ontologies are arguably their inference potential and inherent explainability \cite{allemang_and_hendler_2011, kabashkin_2025}.
Their formal foundations enable us to expand source models by deriving initially unavailable assertions while simultaneously preserving transparency.
To power RDT decision-making heuristics (Section~\ref{contributions}), we thus employ an ontology to serve as the semantic knowledge base, effectively making it a metamodel for describing TT components and user specifications.

To serve as an appropriate knowledge base, the ontology needs to include concepts that can semantically describe a system's elements, such as properties, sensors, and actuators, as well as its desired conditions.
There are at least two well-established candidates which we recognize to be sufficiently close for these purposes: W3C and OGC's Semantic Sensor Network (SSN) \cite{w3c_and_ogc_2017} and ETSI's Smart Applications REFerence (SAREF) \cite{etsi}, both of which are IoT-based ontologies.
Of the two, we chose SSN due to its wide acceptance \cite{compagnucci_et_al_2023}, sufficient coverage of the required concepts, and relatively small size.\footnote{We acknowledge this choice was not made as a result of a systematic evaluation, making SAREF (and its extensions) an equally suitable candidate.}

\begin{figure}[tbp]
    \centering
    \includegraphics[width=1\linewidth]{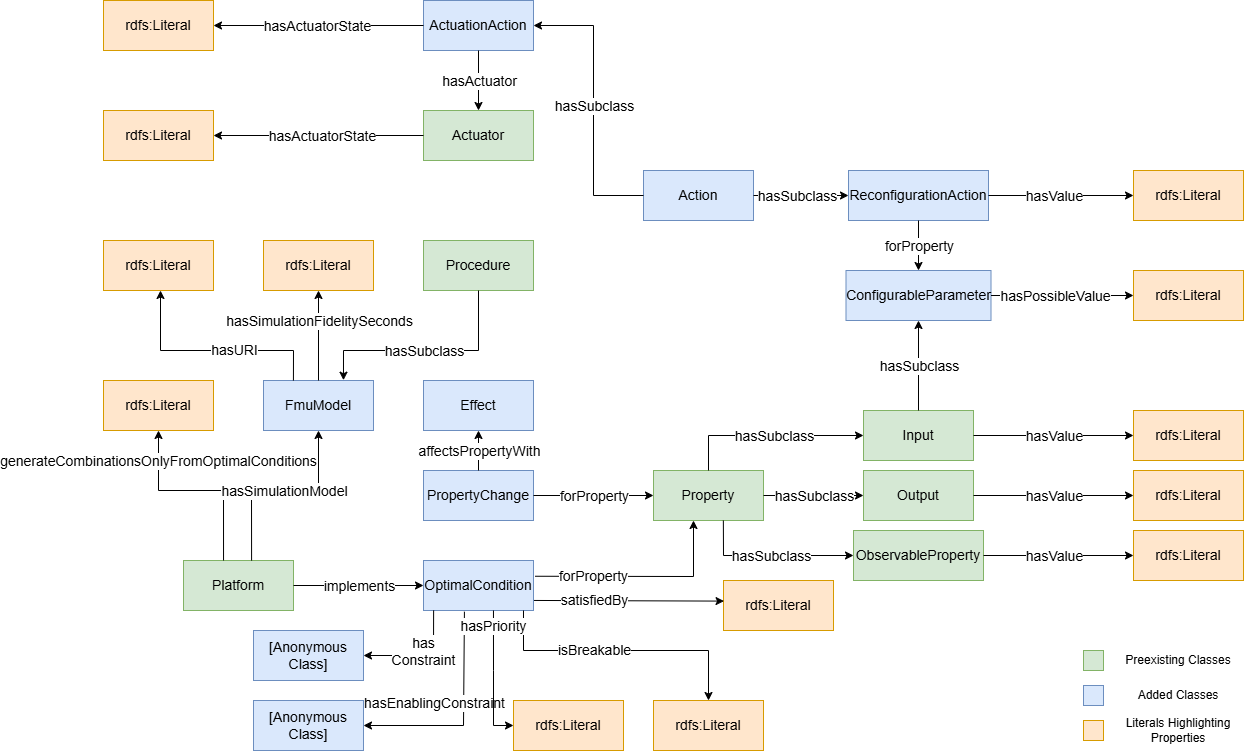}
    \caption{The graph shows all of the added or updated classes and their relations. The literals highlight the newly added data properties.}
    \label{fig:ontology_class_diagram}
\end{figure}

Although SSN already contains concepts like \textit{ssn:Property}, \textit{sosa:Sensor}, and \textit{sosa:Actuator}, additional ones are necessary for RDT decision tree heuristics to work.
We thus extend the ontology by adding new class, object property, and data property definitions, as well as by updating some preexisting class definitions with new statements (triples).
Note that our lightweight augmentations only consist of additions; none of the triples forming any preexisting class definitions have been altered or removed.
As such, the underlying SSN ontology remains unchanged.
Several of the newly added classes essential to enabling RDTs are presented below, whereas the exhaustive collection of added and augmented concepts is shown in Figure~\ref{fig:ontology_class_diagram}.

\subsubsection{\textit{:OptimalConditions}}
\label{optimal_conditions}
\textit{:OptimalConditions}\footnote{We use the empty RDF prefix ``:'' to denote elements specific to our ontological definitions.} define TT conditions considered optimal by the user.
More specifically, much like the value constraints used in MPC (Section~\ref{model_predictive_control}), an \textit{:OptimalCondition} represents a constraint on one or between multiple properties of a TT.
In other words, users express their desired system specifications through \textit{:OptimalConditions}.
The definition of an \textit{:OptimalCondition} is given in Manchester OWL Syntax~\cite{w3c_manchester_owl_syntax} in Listing~\ref{lst:optimal_condition_definition}.
\begin{lstlisting}[label={lst:optimal_condition_definition}, caption={The definition of an \textit{:OptimalCondition} in Manchester Syntax.}]
OptimalCondition
    forProperty exactly 1 Property
    forEnablingProperty exactly 1 Property
    hasConstraint exactly 1 ((comparisonOperator min 1 Property))
    hasEnablingConstraint max 1 ((comparisonOperator min 1 Property))
\end{lstlisting}
The class expression on line 2 states that an \textit{:OptimalCondition} must reference and thus hold for exactly one \textit{ssn:Property}\footnote{Our proof-of-concept implementation technically supports \textit{ssn:Properties} of all RDF types handled by Apache Jena's \cite{apache_software_foundation} equality check functionality (Section~\ref{logical_inference}).}.
Next, following the \texttt{TRIGGER-CONDITION-ACTION} \cite{openhab, soares_et_al_2021, yigal_et_al_2024} rule paradigm, we include a reference to an enabling property (e.g., motion sensor value) on line 3 and two types of conditions for maximum expressivity on lines 4 and 5: firstly, the expression on line 4 represents conditions closely related to those covered by the \texttt{TRIGGER} portion of a traditional rule, albeit with an important distinction.
When developing rules, users would define their \texttt{TRIGGER} portions as \textit{undesirable} conditions that required mitigations.
In contrast, this expression defines the \textit{desirable} conditions, thus aligning with actual user specifications.
Secondly, the expression on line 5 represents conditions that enable the \textit{:OptimalCondition} itself, which allows users to define conditional specifications and potentially retain an element of manual control.
For example, managing comfortable levels of room temperature becomes unnecessary when the inhabitants are away for a prolonged amount of time.
As a result, the expression essentially defines the \texttt{CONDITION} portion of a traditional rule.
\vsnote{Technically we're capturing the pattern in OCs such as: $\mathrm{oc1: COND \to (19 <= temp <= 22)}$. Maybe we can use this to justify user-friendliness, since apparently it ain't strictly necessary?}

While Manchester OWL Syntax allows for expressions combining comparison operators with constants, we allow \textit{:OptimalCondition} constraints to reference \textit{ssn:Properties}.
Doing so creates at least two advantages: firstly, it adds more expressive power by enabling users to specify their desired value ranges in terms of potentially variable property values.
Secondly, even if users wish to define their desired ranges in terms of constant values, defining them as \textit{ssn:Properties} allows for easier instance model management and potential coupling with external systems by avoiding so-called \textit{magic numbers}.

Lastly, to provide a contextual example for office room M370, we define two \textit{:OptimalCondition} individuals for regulating temperature and humidity, as shown in Listings~\ref{lst:m370_optimal_condition_temperature} and~\ref{lst:m370_optimal_condition_humidity}.

\begin{lstlisting}[label={lst:m370_optimal_condition_temperature}, caption={An \textit{:OptimalCondition} individual for \texttt{"If motion sensor activated, keep M370's room temperature between 18°C and 22°C at all times".} Here, \textit{:OfficeTemperatureLowerLimit} = 18, \textit{:OfficeTemperatureUpperLimit} = 22, and \textit{:MotionActivated} = true. \textit{:RoomTemperature} and \textit{:Motion} are properties measured by a sensor.}]
OfficeTemperatureOptimalCondition
    hasConstraint exactly 1 ((greaterThan value OfficeTemperatureLowerLimit) and
        (lessThanOrEqualTo value OfficeTemperatureUpperLimit))
    hasEnablingConstraint exactly 1 (equalTo value MotionActivated)
    forProperty RoomTemperature
    forEnablingProperty Motion
\end{lstlisting}
\begin{lstlisting}[label={lst:m370_optimal_condition_humidity}, caption={An \textit{:OptimalCondition} individual for \texttt{"Keep M370's room humidity between 30\% and 50\%"}. Here, \textit{:OfficeHumidityLowerLimit} = 30 and \textit{:OfficeHumidityUpperLimit} = 50. \textit{:RoomHumidity} is a measured property.}]
OfficeHumidityOptimalCondition
    hasConstraint exactly 1 ((greaterThanOrEqualTo value OfficeHumidityLowerLimit) and
        (lessThanOrEqualTo value OfficeHumidityUpperLimit))
    forProperty RoomHumidity
\end{lstlisting}

\paragraph{Optimization Annotations}
\label{optimization_annotations}
To allow the RDTs to not only keep \textit{:OptimalConditions} satisfied but to also simultaneously optimize for given \textit{ssn:Properties} (in line with the problem described in Section~\ref{problem_domain}), we introduce the \textit{:minimizes} and \textit{:maximizes} object properties as a user specification extension.
These object properties represent desired \textit{ssn:Property} value minimizations and maximizations, respectively, and thus serve as a useful annotation during the ensuing decision ranking and selection process (discussed in Section~\ref{ranking_and_selection_algorithm}).


\subsubsection{\textit{:Effects} and \textit{:PropertyChanges}}
\label{effects_and_property_changes}
As mentioned in Section~\ref{contributions}, one of the secondary contributions and features of an RDT is its ability to apply heuristics to its decision-making process.
To demonstrate this feature, we will use a working example of a \textit{reactive} approach to actuator control.
With reactive heuristics, an RDT performs a trade-off: by only considering actions that directly mitigate already violated \textit{:OptimalConditions}, it reduces the size of its decision trees at the cost of sacrificing its proactive control ability.

To automatically infer mitigations for violated \textit{:OptimalConditions}, the semantic knowledge base needs information about not only what properties actuators affect but also \textit{how} they affect them.
To describe the kind of effect an actuator may have on a property, we introduce the \textit{:Effect} class.
The class itself has no special definition but rather serves as a container for two individuals, namely, \textit{:ValueIncrease} and \textit{:ValueDecrease}, which describe property value changes in a simple, qualitative manner.
Although these could be defined as simple object properties, a class offers more flexibility for potentially adding more complex and precise \textit{:Effect} types in the future (further discussed in Section~\ref{discussion}).

To ontologically connect \textit{sosa:Actuators} and \textit{:ConfigurableParameters} to their respective \textit{ssn:Properties} and \textit{:Effects}, we introduce the \textit{:PropertyChange} class, whose definition is shown in Listing~\ref{lst:property_change_definition}.
\begin{lstlisting}[label={lst:property_change_definition}, caption={The definition of a \textit{:PropertyChange} in Manchester OWL Syntax.}]
PropertyChange
    forProperty exactly 1 Property
    affectsPropertyWith exactly 1 Effect
\end{lstlisting}
A \textit{:PropertyChange} acts as a wrapper for references to both a property and the effect applied on it.
This is necessary in cases of actuators simultaneously affecting multiple properties in different ways.
For example, an air conditioning unit may affect room temperature with a \textit{:ValueDecrease} while also affecting energy consumption with a \textit{:ValueIncrease}.
To that end, connecting \textit{sosa:Actuators} (and \textit{:ConfigurableParameters}) to \textit{ssn:Properties} through \textit{:PropertyChanges} preserves the relation between those \textit{ssn:Properties} and the \textit{:Effects} applied on them.

Listings \ref{lst:m370_heater}, \ref{lst:m370_room_temperature_increase}, and \ref{lst:m370_energy_consumption_increase} demonstrate how a heater in office M370 affects its room temperature and energy consumption.
\begin{lstlisting}[label={lst:m370_heater}, caption={The definition of a heater with three states in office M370.}]
Heater
    Actuator
    enacts RoomTemperatureIncrease
    enacts EnergyConsumptionIncrease
    hasActuatorState "0"^^xsd:int
    hasActuatorState "1"^^xsd:int
    hasActuatorState "2"^^xsd:int
\end{lstlisting}
\begin{lstlisting}[label={lst:m370_room_temperature_increase}, caption={The definition of a temperature property change in office M370.}]
RoomTemperatureIncrease
    PropertyChange
    affectsPropertyWith ValueIncrease
    forProperty RoomTemperature
\end{lstlisting}
\begin{lstlisting}[label={lst:m370_energy_consumption_increase}, caption={The definition of an energy consumption property change in office M370.}]
EnergyConsumptionIncrease
    PropertyChange
    affectsPropertyWith ValueIncrease
    forProperty EnergyConsumption
\end{lstlisting}

\subsubsection{\textit{:Actions}}
\label{actions}
To represent the atomic changes RDTs make on their TTs as a result of their decisions, we introduce the \textit{:Action} class as well as its two disjoint subclasses, \textit{:ActuationAction} and \textit{:ReconfigurationAction}, whose definitions are shown in Listings~\ref{lst:actuation_action_definition}~and~\ref{lst:reconfiguration_action_definition}, respectively.
\begin{lstlisting}[label={lst:actuation_action_definition}, caption={The definition of an \textit{:ActuationAction} in Manchester OWL Syntax. \textit{:ActuationActions} handle actions related to \textit{sosa:Actuators}.}]
ActuationAction
    Action
    hasActuator exactly 1 Actuator
    hasActuatorState exactly 1 rdfs:Literal
\end{lstlisting}
\begin{lstlisting}[label={lst:reconfiguration_action_definition}, caption={The definition of a \textit{:ReconfigurationAction} in Manchester OWL Syntax. \textit{:ReconfigurationActions} handle actions related to \textit{:ConfigurableParameters}.}]
ReconfigurationAction
    Action
    forProperty exactly 1 ConfigurableParameter
    hasValue exactly 1 rdfs:Literal
\end{lstlisting}
\textit{:Actions} are simple wrappers referencing a TT component and its potential new state or value.
If an RDT chooses to execute an \textit{:Action} as part of its decision, then the referenced component's own state or value is set to the new one.
For example, to set M370 \textit{:Heater's} state to ``2'', the RDT would execute the \textit{:ActuationAction} shown in Listing~\ref{lst:actuation_action_example}.
\begin{lstlisting}[label={lst:actuation_action_example}, caption={An example of an \textit{:ActuationAction} for setting M370 \textit{:Heater's} state to ``2''.}]
ExampleActuationAction
    ActuationAction
    hasActuator Heater
    hasActuatorState "2"^^xsd:int
\end{lstlisting}

It is important to note that unlike the previously introduced concepts, \textit{:Actions} are not provided by the user and are thus not a part of the \textit{core} instance model data.
They are instead inferred from that data and subsequently form the \textit{expanded} (inferred) instance model, as discussed in Sections~\ref{inference_rules}~and~\ref{mape-k_loop}.

\subsection{Logical Inference}
\label{logical_inference}
A crucial element of an RDT's decision-making is the utilization of logical inference.
It lets us exploit the semantics of the knowledge base to dynamically expand a TT instance model based on a combination of the static and dynamic (observed) data within it.
Additional data is commonly asserted through the execution of declarative inference rules consisting of propositional clauses of the form
\[ antecedent \longrightarrow consequent \]
where the \( antecedent \) represents a rule's body and the \( consequent \) its head \cite{kowalski_1974}.
In ontological inference, the body of a rule contains one or more assertions (triples) to match on, while its head consists of the resulting assertions to produce.
New assertions can ultimately be inferred once all triples from a rule's body are found to be true (exist as stated) in the dataset being reasoned over.

Inference rule parsing and execution is handled by semantic reasoners or so-called inference engines.
Due to the nondeterministic nature of logic programming \cite{genesereth_and_ginsberg_1985, kowalski_1974}, rule execution order cannot be guaranteed and should thus not be expected \cite{the_apache_software_foundation_inference}.
Reasoners will therefore execute a rule irrespective of others and only when both of the following conditions are true: (\textit{i}) all of the triples of the rule's body match those of the dataset and (\textit{ii}) there are (potentially new) triples in the dataset the rule can but has not yet matched on.
To power logical inference in our own proof-of-concept RDT, we employ Apache Jena's \cite{apache_software_foundation} open-source inference engine due to its long-standing establishment and use, extensive support for RDF/OWL ontological models, and intuitive rule syntax.

\subsubsection{Inference Rules}
\label{inference_rules}

\begin{lstlisting}[label={lst:inference_rule_example}, caption={An example of an inference rule used in the later stages of the algorithm. Rule 6a finds \textit{:PropertyChanges} and the \textit{sosa:Actuators} that cause them, and generates corresponding \textit{:Actions} (with debugging output).}]
[rule6a:   (?thing :withPropertyChange ?propertyChange)
           (?actuator meta:enacts ?propertyChange)
           (?actuator rdf:type sosa:Actuator)
           (?actuator meta:hasActuatorState ?actuatorState)
           uriConcat(?actuator, "_", ?actuatorState, ?actionName)
           ->  (?actionName rdf:type meta:ActuationAction)
               (?actionName meta:hasActuator ?actuator)
               (?actionName meta:hasActuatorState ?actuatorState)
               print("Generated ActuationAction", ?actionName)
]
\end{lstlisting}

Although reasoners like Protege's \cite{musen_and_protege_2015} HermiT \cite{glimm_hermit_2014} correctly infer functional relations, class subsumptions, and much more, they cannot infer necessary \textit{:Actions} as a result of violated \textit{:OptimalConditions}.
Because OWL is intended as a general ontology language, it simply does (and should) not offer the constructs to describe such domain-specific semantic relationships.
Therefore, to complete the semantic definitions of our ontological entities (Section~\ref{semantic_knowledge_base}), and subsequently enable a more domain-specific inference mechanism, we must rely on custom inference rules.

Our inference rules (Listing \ref{lst:inference_rule_example}) are organized such that they effectively form an algorithm based on the following semantic connection (Section~\ref{semantic_knowledge_base}):
\[ \textit{:OptimalCondition} \rightarrow \textit{ssn:Property} \rightarrow \textit{:PropertyChange} \rightarrow \textit{sosa:Actuator} \rightarrow \textit{:Action} \]
Given an ontologically conforming TT instance model\footnote{Ontological model-checking may be performed with frameworks like SHACL \cite{w3c_shacl_2017}.}, the algorithm proceeds in one of two ways: by default (core RDT functionality), the rules match on all \textit{sosa:Actuators} found in the instance model, construct \textit{:Actions} from each one's set of states, and produce valid \textit{:Action} combinations, with each representing a possible RDT decision.
Otherwise, provided the RDT uses reactive decision-making heuristics (secondary RDT functionality), the inference process begins with the specified \textit{:OptimalConditions}.
For example, given an \textit{:OptimalCondition} with its referenced \textit{ssn:Property's} lower bound violated, the RDT infers that the relevant mitigation must consist of a \textit{:PropertyChange} on the same \textit{ssn:Property} with a \textit{:ValueIncrease} \textit{:Effect}.
In the case of physical components, it can then find all of the TT's \textit{sosa:Actuators} that cause the required \textit{:Effect} on the corresponding \textit{ssn:Property} and consequently infer all possible \textit{:Actions} and thus \textit{:Action} combinations (decisions) to enact.

\begin{figure}[tbp]
    \centering
    \includegraphics[width=1\linewidth]{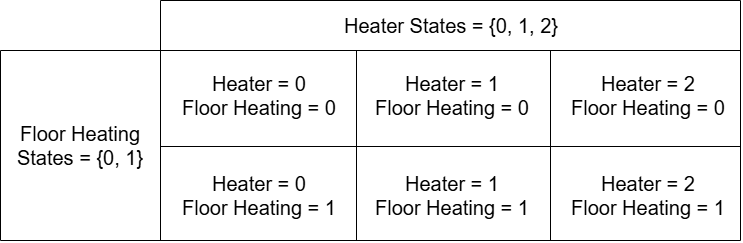}
    \caption{A table of inferred RDT decisions (\textit{:Action} combinations) as the Cartesian product of M370's heating \textit{sosa:Actuator} states.}
    \label{fig:inferred_action_combinations}
\end{figure}

To provide an example in the context of office room M370, we assume that an RDT uses reactive decision-making heuristics and that \textit{:RoomTemperature's} value is low enough to violate \textit{:OfficeTemperatureOptimalCondition} (Listing~\ref{lst:m370_optimal_condition_temperature}).
Given M370's heating actuators (Figure~\ref{fig:office_room_m370}), the RDT will infer six potential decisions in total, as shown in Figure~\ref{fig:inferred_action_combinations}.

\subsection{Simulations}
\label{simulations}
RDTs rely on simulations to explore their potential decisions' impacts through hypothetical scenarios.
Somewhat analogously to MPC, simulating decisions before enacting them allows an RDT to compare and evaluate its predicted results, thereby increasing its accuracy and enhancing its decision-making process.
The simulations in our proof-of-concept implementation are powered by the Functional Mock-up Interface (FMI) \cite{blockwitz_et_al_2012, modelica_association} standard.
This widely-adopted standard facilitates exchanges of simulation models in the form of high-performance, executable binaries known as Functional Mock-up Units (FMUs).

\begin{figure}[tbp]
    \centering
    \includegraphics[width=1\linewidth]{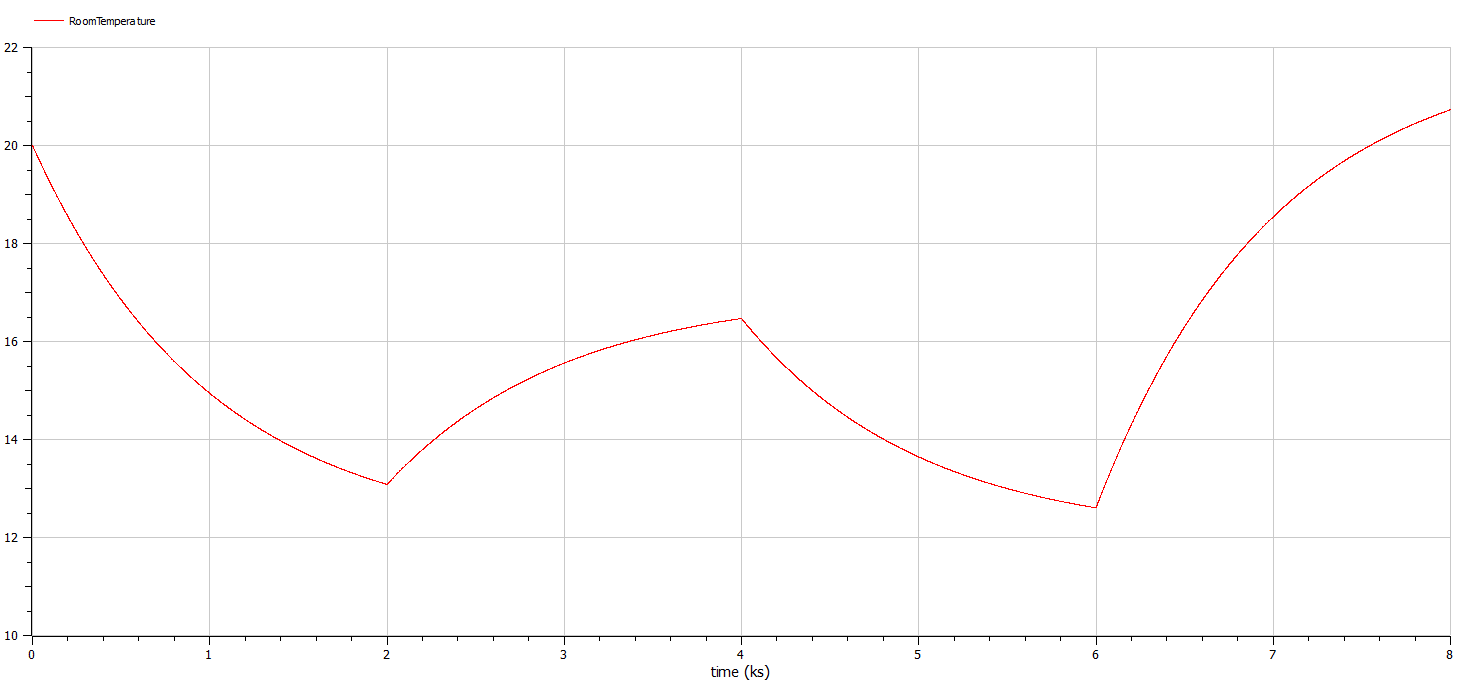}
    \caption{The simulated temperature curve of office room M370 over 8000s with a starting value of 20\textdegree C and its \textit{:Heater} state set to ``0'' at \( t=0s \), ``1'' at \( t=2000s \), ``0'' at \( t=4000s \), and ``2'' at \( t=6000s \). The simulation was conducted and visualized with OpenModelica and OMEdit \cite{fritzson_et_al_2020}.}
    \label{fig:m370_fmu_temperature_curve}
\end{figure}

Because an actuator may simultaneously affect multiple properties, and because multiple actuators may simultaneously affect the same property, it is necessary to build FMUs that model the relationships between all interacting elements of a TT.
Our framework thus requires that an FMU corresponding to a specific ontological instance model include all of its actuators and directly affected properties as parameters of its own model.
To that end, office room M370's FMU implementation contains rudimentary differential equations representing the rates of change of temperature and humidity values based on respective actuations.\footnote{Note that these equations are not accurate representations of physical phenomena but serve as adequate placeholders for our proof of concept.}
An example of a simulated temperature curve under varying states of actuation is shown in Figure~\ref{fig:m370_fmu_temperature_curve}.

\subsection{Multi-Level Hypotheticals}
\label{multi_level_hypotheticals}
By their core functionality, RDTs simulate hypothetical decisions over a single time period representing the current `Monitor, Analyze, Plan, and Execute over Knowledge' (MAPE-K) cycle (Section~\ref{mape-k_loop}).
As such, decisions are always taken with respect to enacting the optimal \textit{:Action} combinations for that cycle alone.
For example, given M370's \textit{:OfficeTemperatureOptimalCondition} and a simultaneous specification for minimizing energy bills, an RDT may be faced with multiple decision options.
In case all of those options restore the \textit{:OptimalCondition}, the RDT will choose the one keeping its energy bills minimal in the current cycle.

However, assuming unchanging user specifications, always exclusively picking the decision optimal for that cycle alone might be suboptimal in the long run.
Instead, similarly to MPC \cite{nikolaou_model_2001}, we enable RDTs to extend their prediction horizons to explore hypothetical scenarios spanning multiple upcoming cycles.
As a result, they produce and handle decision trees much like those found in mixed-integer MPC (Section~\ref{mixed-integer_model_predictive_control}).
Each unique path (root to leaf) on these decision trees represents a unique sequence of a number of RDT decisions for the matching number of upcoming cycles. 
Note that by default (without updating static TT component information), RDTs construct identical trees each time due to each upcoming cycle's (tree level's) decisions always including all of the TT's actuators (and configurable parameters).
This results in symmetric trees with an equal number of branches spanning from each node on each given level.
Conversely, with reactive decision-making heuristics, RDTs produce asymmetric trees due to each upcoming cycle's (tree level's) decisions depending on the (simulated) results of their ancestors.

Naturally, depending on the length of the prediction horizon (and the heuristics used), constructing such decision trees may lead to combinatorial blowups.
This directly results in increased simulation and result evaluation times, thus also restricting the minimum duration of a MAPE-K cycle.
Furthermore, due to potential modeling inaccuracies, simulating hypothetical scenarios multiple cycles in advance bears inherent, accumulated uncertainty.
Even if one tolerates longer computations, initially small deviations (modeling inaccuracies) between simulations and the real world only grow larger with each additional cycle included, subsequently imposing practical limits on how long a useful prediction horizon may be.

\subsubsection{Cycle Durations}
\label{cycle_durations}
\begin{figure}[tbph]
  \begin{minipage}[b]{0.7\linewidth}
    \includegraphics[width=1\linewidth]{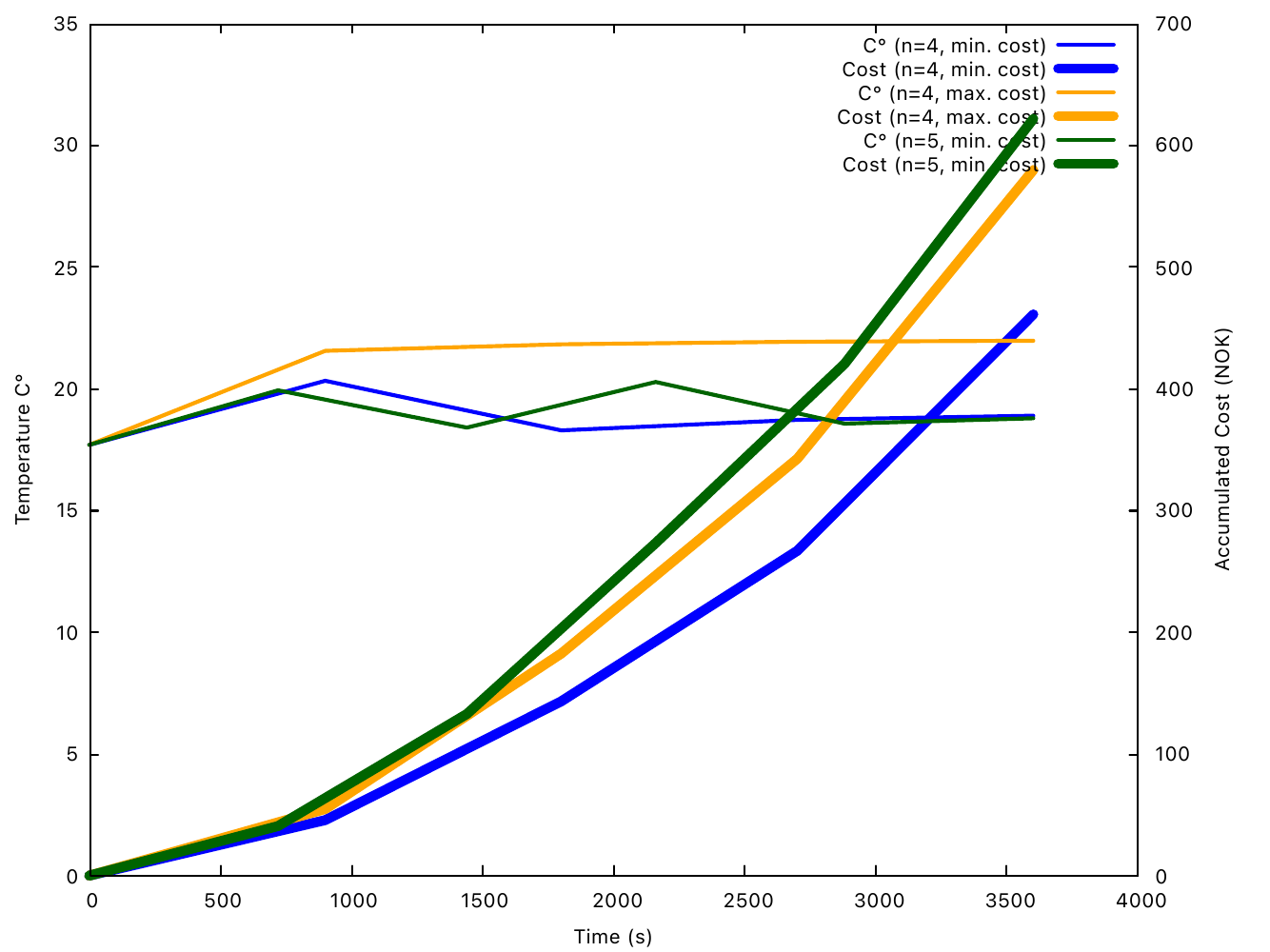}
    \caption{A demonstration of different cycle durations having different impacts on the accumulated energy prices.}
    \label{fig:multi-level_hypothetical_comparison}
\end{minipage}\quad
\begin{minipage}[b]{0.25\linewidth}
  \begin{lstlisting}[numbers=none]
;;; fakepool.tsv
price
2.323
2.026
1.671
1.617
1.598
1.586
1.479
1.479
1.479
1.484
1.62
...
\end{lstlisting}
\vspace*{1cm}
\caption{Historic price data used in the simulation.}
\label{fig:fakepool}
\end{minipage}
\end{figure}

To illustrate the differences in cycle durations relating to minimizing energy bills for office room M370, we simulated three different control scenarios over an hour (\num{3600}s) (Figure~\ref{fig:multi-level_hypothetical_comparison}).
For each scenario, the chart plots the temperature in the room (thin lines) and the respective accumulated cost (thick lines).
We used a historic snapshot of the Nord Pool electricity spot price split into 15-minute intervals (Figure~\ref{fig:fakepool}).
The first two scenarios (blue and orange lines, ``\texttt{n = 4}'') split the entire 1-hour interval into four 15-minute cycles, whereas the final scenario splits it into five (green, ``\texttt{n = 5}'') of correspondingly shorter duration.

Although the execution of a given cycle may last up to its assigned interval duration (e.g., 15 minutes), each one executes in less than that time (assuming a prediction horizon of 3, approximately 13 seconds).
Thus, between enacting changes and conducting observations in the next cycle, there is typically a pause during which actuators or configurable parameters remain in their previously assigned states or values.

The two scenarios annotated with ``\texttt{min. cost}'' (blue and green) pick decisions that keep the room climate comfortable (as per our \textit{:OptimalConditions}) and minimize the accumulated cost.
For the other scenario (``\texttt{max. cost}'', orange), we removed the minimization annotation from the model and selectively plotted the most expensive but comfortable solution.

Comparing the first two (``\texttt{n = 4}''), we see that the minimization picks a plan that results in lower cost and slightly lower room temperature.
The difference between the first (blue) and the last (green) scenario shows the somewhat (in hindsight) expected effect that shorter scenario cycles increase precision.
The reason behind the more fine-grained scenario ending with a higher cost is that, due to the nature of the price data (in 15-minute intervals) coinciding with the cycle duration, we were in fact underestimating the cost of electricity, as our price data (Figure \ref{fig:fakepool}) has a generally falling trend in the relevant interval.
With its shorter interval, the green scenario generally picks up the correct price more frequently and also produces a smoother temperature curve.
Due to the falling price trend, the final scenario also slightly delays heating until \num{2160} seconds.

\subsection{Case-Based Reasoning}
\label{case_based_reasoning}
As explained in Sections \ref{simulations} and \ref{multi_level_hypotheticals}, simulations are an integral part of the RDT decision-making process.
Although necessary for decision result predictions, their execution and subsequent result evaluation impose a potentially substantial computational cost (Section~\ref{multi_level_hypotheticals}).
Over a span of many cycles, however, there is a chance an RDT repeatedly observes (nearly) identical TT conditions to those encountered before.
Provided all models (including all user specifications) remain unchanged, upon encountering such conditions, the DT will naturally rerun the same simulations and consequently reenact the same decisions.

To capitalize on already incurred computational cost, we enable RDTs to employ case-based reasoning (CBR) \cite{aamodt_and_plaza_1994}.
CBR utilizes knowledge from previously-encountered scenarios (cases) to handle not-yet-encountered but similar ones.
To that end, an RDT could do the same by saving information about the TT conditions it observes and the respective successful decisions it takes.
Upon encountering conditions similar enough to those in a saved case (discussed in Section~\ref{analyze_and_plan}), it would simply execute its previously derived decision, thereby avoiding unnecessary inference, simulations, and result evaluations.

Our own proof-of-concept RDT uses case models containing the following components: (\textit{i}) TT conditions (\textit{ssn:Properties}), (\textit{ii}) \textit{:OptimalConditions}, (\textit{iii}) the length of the prediction horizon (number of cycles), (\textit{iv}) interval length (e.g., 15 minutes), (\textit{v}) decision index (tree level of decision), and (\textit{vi}) the decision (\textit{:Action} combination) itself.
We specify that matching must occur not just on component \textit{i} but also on components \textit{ii} -- \textit{v} since, even if the observed TT conditions match, making the same decision is only rational if user specifications and other configurations do too.
Component \textit{iv} represents the interval length and thus also the virtual duration of a simulation (full in-simulation time), whereas component \textit{v} represents the ordinal number of the case's decision (and thus the case itself) in the context of a potential multi-cycle decision execution.
Lastly, component \textit{vi} represents the decision containing both the \textit{:Actions} to enact and the expected property values as a result thereof, which are subsequently used as a form of verification for successful case application \cite{bolender_et_al_2021, michael_et_al_2024}.

With more conditions encountered and cases generated, an RDT's average cost of computation per cycle is expected to gradually drop, ultimately exchanging computation for memory.
Moreover, as a consequence of a growing case base, we effectively automate the generation of a traditional, rule-based decision-making model.
For example, it is clear that case component information can be mapped (via, e.g., a model transformation) onto all of the portions of a \texttt{TRIGGER-CONDITION-ACTION} \cite{openhab, soares_et_al_2021, yigal_et_al_2024} rule.

\subsection{MAPE-K Loop}
\label{mape-k_loop}

\begin{figure}[tbp]
    \centering
    \includegraphics[width=1\linewidth]{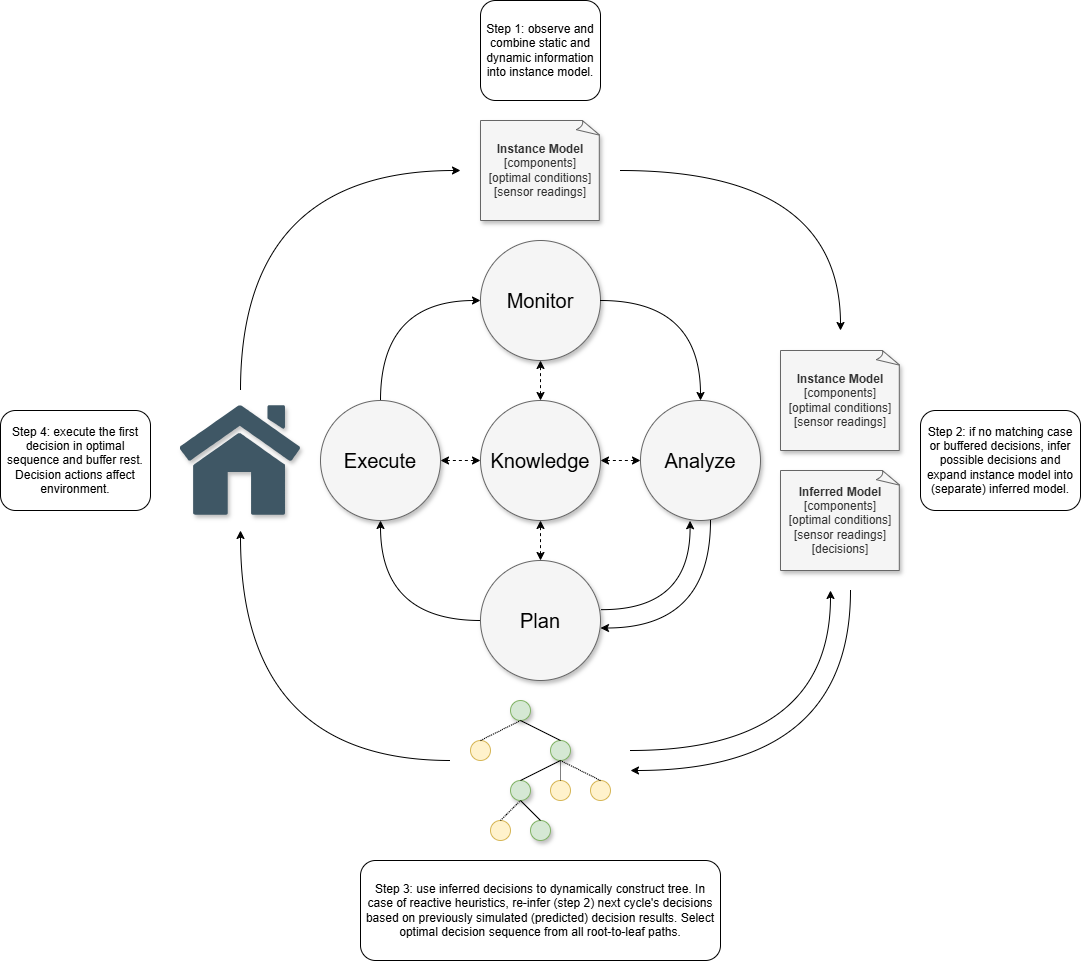}
    \caption{}
    \label{fig:mape_cycle}
\end{figure}

For RDTs to conduct observations and enact their respective decision-making, we require that they incorporate aspects of autonomic computing.
Due to its ubiquity and suitability for DTs \cite{feng_et_al_2022}, we utilize the `Monitor, Analyze, Plan, and Execute over Knowledge' (MAPE-K) architecture \cite{kephart_and_chess_2003}.
A standard MAPE-K cycle consists of four distinct and ordered phases, each one exchanging information with a knowledge base representing the Knowledge component.
The cycle begins with the Monitor phase gathering data, such as sensor readings and component details, about the managed system.
Next, the Analyze phase realizes execution options based on what is observed about the managed system and the information in the knowledge base.
With the provided options, the Plan phase typically utilizes simulations and/or planning algorithms to explore hypothetical scenarios and produce a viable execution plan.
Finally, the Execution phase implements this plan by enacting changes onto the managed system.
All four phases are typically executed as quickly as possible to minimize cycle durations and allow for more frequent decision-making.

We implement the MAPE-K loop and the rest of our proof-of-concept RDT in .NET and thus use the dotNetRDF \cite{vesse_et_al} and Femyou \cite{azeau} libraries for managing ontological concepts and FMUs, respectively.
To interact with the knowledge base (a Turtle \cite{beckett_et_al_2014} file), the RDT relies on querying with SPARQL \cite{w3c_sparql_2013}.

In the following subsections, we outline how each of the four MAPE phases manifests within our RDT as well as the slight alteration our implementation requires in their execution order.
The whole process is illustrated in Figure~\ref{fig:mape_cycle}.
Lastly, we demonstrate a concrete example of a MAPE-K cycle as part of the RDT managing (virtual) office room M370.

\subsubsection{Monitor}
\label{monitor}
Since almost all MAPE phases in our implementation exchange information with the semantic knowledge base, the Monitor phase begins by constructing a cache of relevant data to reduce further querying.
Cache construction begins by querying for the following static TT information: (\textit{i}) the properties, sensors, soft sensors \cite{jiang_et_al_2021}, configurable parameters, and actuators the TT consists of, (\textit{ii}) current user specifications in the form of \textit{:OptimalConditions}, and  (\textit{iii}) how the discovered soft sensors relate to certain properties (inputs/outputs) and build on other soft sensors to make \textit{soft sensor trees}.

Because soft sensors are essentially software (algorithms) \cite{jiang_et_al_2021}, it is easily conceivable that they use other soft sensors' outputs as their own inputs.
As such, users may include soft sensors building on other soft sensors as part of their TTs.
To facilitate the appropriate management and use of properties and their values relating to such \textit{soft sensor trees} (e.g., averages calculated from multiple other sensors\footnote{These may be hard or soft sensors.}), our RDT implementation deliberately keeps track of their structure.

During construction, all queried information is subsequently used in acquiring additional dynamic data about the TT's current conditions (sensor observations and property values) and state (actuator states and/or configurable parameter values).
For example, when querying the knowledge base for the TT's observable properties, the RDT also queries for the sensors observing them.
Unique sensor information (their URIs) enables the framework to invoke the corresponding, user-provided connection adapters to retrieve the respective sensors' current readings and add values to the properties in the cache.
With a completed cache, the Monitor phase captures all information pertinent to the TT in a given cycle.

Observed property value data is ultimately written into the knowledge base to prepare it for the RDT's inference process (in case of reactive heuristics used), thereby effectively combining the static and dynamic TT models.
Due to the Monitor phase re-querying the knowledge base at the start of every cycle, it not only keeps the RDT updated with more current dynamic data but also enables dynamic changes to both static TT information (e.g., new components or more actuator states) and user specifications (e.g., changing \textit{:OptimalConditions}).

\subsubsection{Analyze and Plan}
\label{analyze_and_plan}
Before the Analyze and Plan phases, our implementation performs checks relating to multi-level hypotheticals (Section~\ref{multi_level_hypotheticals}) and CBR (Section~\ref{case_based_reasoning}) functionality.
Firstly, if the RDT is configured to simulate for and generate multi-cycle decisions (decisions covering the length of the prediction horizon), then there is a possibility some of them are already buffered (from a previous cycle) and yet to be executed.
If there is at least one remaining buffered decision to be executed, and the TT is within the user-configured deviation tolerance of the predicted conditions resulting from the last decision's execution (in the previous cycle), then the buffered decision is chosen, and the current cycle's Analyze and Plan phases are skipped.
If there are no remaining buffered decisions and CBR functionality is enabled, the framework queries the case database with the TT's current conditions to find a match(es) for the configured prediction horizon.
Given a match, the case's precomputed decision is chosen, and the Analyze and Plan phases are likewise skipped.
If there are no remaining buffered decisions or matching cases to execute, or if the observed TT conditions deviate outside the user-configured tolerance value from the last cycle's predictions, the Analyze and Plan phases are initiated.

\begin{figure}[tbp]
    \centering
    \includegraphics[width=1\linewidth]{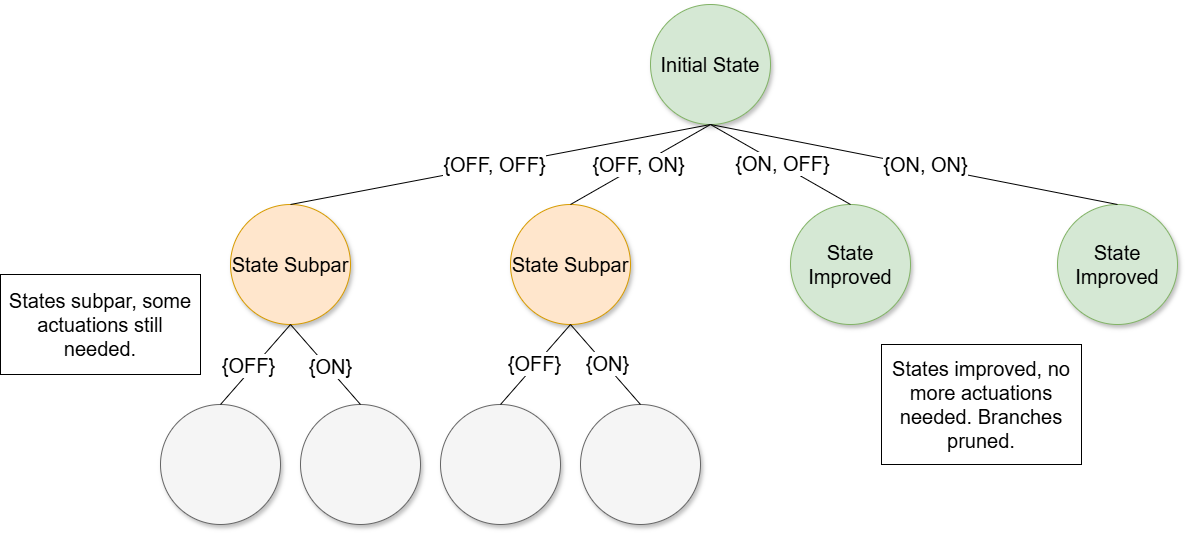}
    \caption{Due to the dynamic nature of decision tree construction, branches can effectively be pruned away based on the results of a given simulation (tree node) \cite{spajic_and_stolz_2026}.}
    \label{fig:decision_tree_pruning_example}
\end{figure}

In terms of the MAPE-K loop, the RDT's inference process is treated as the Analyze phase, while its simulations and subsequent result evaluations form the Plan phase.
The inference step expands the initial instance model into a separate inferred model now also containing possible decisions.
Keeping the two separate in the knowledge base ensures the original instance model stays clean and prevents indefinite growth and pollution of the inferred one over many MAPE-K cycles.
After inferring possible decisions, the RDT handles simulation generation in a dynamic streaming fashion to allow any potential heuristics to prune the decision tree (Figure~\ref{fig:decision_tree_pruning_example}).
For every successfully inferred decision, a simulation is generated and immediately executed for its predictive results.
In case the RDT uses multi-level hypothetical functionality (Section~\ref{multi_level_hypotheticals}), these results become especially relevant in conjunction with reactive decision-making heuristics as they directly affect the number of possible decisions in subsequent cycles (decision tree levels).
Once simulation results are stored in the knowledge base, the inference process starts again to infer possible decisions for the following cycle.\footnote{For performance reasons, if no decision-making heuristics are used, the RDT does not run the inference process again since the previously-generated \textit{:Action} combinations are known to remain the same in the current MAPE-K cycle.}
Altogether, our implementation deviates from the standard MAPE phase running order with Analyze and Plan being more interwoven and potentially executed multiple times in a single cycle.
This process ultimately produces a tree where each path (root to leaf) represents a possible TT future consisting of hypothetical decisions and their outcomes over the prediction horizon.

\paragraph{Ranking and Selection Algorithm (Optimization Function)}
\label{ranking_and_selection_algorithm}
To find the optimal out of all root-to-leaf paths on the decision tree, our implementation utilizes a rudimentary ranking and selection (R\&S) algorithm \cite{hong_et_al_2021, walton_et_al_2024}.
In terms of MPC, this R\&S algorithm forms the optimization function yielding optimal process inputs (actuator states) over a prediction horizon (simulated upcoming cycles) with respect to process value constraints (user-specified \textit{:OptimalConditions}).

Firstly, each path is assigned a score based on how many \textit{:OptimalCondition} bounds all of its decisions satisfy over the prediction horizon, and all paths with the highest score are retained.
Next, every remaining path is assigned a new score based on its predicted property values.
As explained in Section~\ref{optimization_annotations}, the RDT will attempt to optimize property values based on the respective annotations in its knowledge base.
All optimized property values are compared to corresponding property values in other paths, and those paths with the resulting lowest/highest (most optimized) values have their scores incremented for each such property.
For example, if property \( \mathcal{P} \) is to be minimized, \( \mathcal{P} \)'s values are compared across all paths such that those with the lowest \( \mathcal{P} \) (accumulated across all cycles) have their scores incremented.
The paths with the highest score are retained.
In case of multiple paths still remaining, all are considered optimal, one is chosen arbitrarily, and its decisions are used for execution.

\begin{lstlisting}[float={htbp}, caption={Object-oriented expression examples that capture the product of two observables and accumulate over time.}, label={lst:derived}]
var f_energy = new FProp(":EnergyConsumption");
var f_temp = new FProp(":price");
var f_prod = new FBinOpArith(f_energy, f_temp, (x, y) => x * y, name: ":EnergyTimesPrice");
var f_prod_acc = new FAcc<double>(f_prod, name: ":AccumulatedEnergyTimesPrice");
\end{lstlisting}

This scheme is currently hard-coded in the central MAPE-K loop.
To give the user more flexibility and to allow them to specify optimal conditions on derived properties with their problem domain,
they can write derived properties, e.g.,\ based on binary operations and accumulation, on the .NET-level for their model (Listing~\ref{lst:derived}).
As our RDT framework does not have an integrated development environment (IDE) yet, we leave a user-facing editor/scripting language for future work (discussed in Section~\ref{future_work}).
These expressions compute derived values per simulation and allow comparison between them.
Properties with these names must already be declared in the model, and can be used, e.g.,\ in optimal conditions.
This allows the planner to rank paths with more flexibility than a model that relies on the binary satisfaction/dissatisfaction of optimal conditions and property optimizations.
The same mechanism is then used to encode their preference on solving the inherent cost vs. comfort trade-off (identifying solutions on the Pareto-front) through the planner.

Consider the example where it is only possible to fulfill one of two optimal conditions, depending on the \textit{:Actions} chosen.
Here, the user could, through the comparison function, state a preference, making one more important than the other.

\subsubsection{Execute}
\label{execute}
With a determined sequence of decisions, the RDT begins executing the \textit{:Actions} of the first one by setting corresponding actuators and configurable parameters to their new state or value, which they keep until the next cycle's Execution phase (assuming they will change).
Any remaining decisions (in a multi-cycle path) are buffered for subsequent cycles and are only enacted if the predicted TT conditions of the previously executed decision fall within the user-configured deviation tolerance of those observed in the new cycle.
If the newly observed conditions match (all differences are within the tolerance), the Analyze and Plan phases are bypassed, and the buffered decision is immediately executed.
In case of CBR functionality being used, matching conditions would demonstrate the previous decision as a successful one, prompting the RDT to produce and store a new case in its case base.

\subsubsection{M370 Example Demonstration}
\label{m370_example_demonstration}
\begin{lstlisting}[label={lst:m370_example_configuration}, caption={Parts of the framework configuration for twinning office room M370 shown in JSON.}]
"CoordinatorSettings": {
  ...
  "StartInReactiveMode": false,             // reactive decision-making heuristics
  "UseCaseBasedFunctionality": false,       // CBR
  ...
  "CycleDurationSeconds": 900,              // length of MAPE cycle (time between Monitor phases)
  "LookAheadMapekCycles": 3,                // prediction horizon length (in cycles)
  "PropertyValueFuzziness": 2,              // observed value deviation tolerance
                                               (used for all properties)
  ...
}
\end{lstlisting}
\begin{lstlisting}[label={lst:m370_example_conditions}, caption={Starting values of (relevant) \textit{ssn:Properties} in office room M370.}]
RoomTemperature = 17.7
RoomHumidity = 10.2
EnergyConsumption = 0.0
\end{lstlisting}

In the following section, we complete the demonstration of our proof-of-concept RDT by showing how it updates virtual office room M370's instance model over a complete MAPE-K cycle execution.
Listings \ref{lst:m370_example_configuration} and \ref{lst:m370_example_conditions} show parts of the RDT's configuration and (the more relevant) starting TT conditions, respectively.
\vsnote{I wouldn't waste a figure on the values on Lst. 15. For 14, I'm not sure if we need to show JSON, are can just list/explain the 5 parameters in a sentence instead.}\isnote{I have no hard opinions, but it's a nice visual in case they end up playing with it.}

In the Monitor phase, TT components and user specifications (static information) are gathered from the knowledge base, and the starting conditions (dynamic information) are observed via their respective, user-provided sensor connection adapters (Section~\ref{monitor}).
The static and dynamic information is combined and subsequently saved into the knowledge base, as shown by Listing~\ref{lst:m370_example_monitor_update}.

\begin{lstlisting}[label={lst:m370_example_monitor_update}, caption={A snippet of the knowledge base (\texttt{.ttl} file) at the time of a completed Monitor phase execution showing how M370's sensor readings combine with the corresponding \textit{ssn:Properties}.}]
// Current M370 room temperature.
:RoomTemperature rdt:hasValue "17.7"^^xsd:double;
                 a <http://www.w3.org/ns/sosa/ObservableProperty>,
                   owl:NamedIndividual.

// Current M370 room humidity.
:RoomHumidity rdt:hasValue "10.2"^^xsd:double;
              a <http://www.w3.org/ns/sosa/ObservableProperty>,
                owl:NamedIndividual.

// Current M370 energy consumption.
:EnergyConsumption rdt:hasValue "0"^^xsd:double;
                   a <http://www.w3.org/ns/sosa/ObservableProperty>,
                     owl:NamedIndividual.
\end{lstlisting}

Since there are no buffered decisions to check for in the first cycle, the RDT immediately follows up with its Analyze and Plan phases.
Although it is clear that both of M370's \textit{:OptimalConditions} (Section~\ref{optimal_conditions}) have their constraints violated (assuming they are both enabled), the inference process in the Analyze phase ignores them due to the RDT not using reactive decision-making heuristics.
With the TT's \textit{:Heater} having 3 states and its \textit{:FloorHeating} and \textit{:Dehumidifier} each having 2, the RDT produces 12 hypothetical decisions for the current cycle.
Given that the prediction horizon is 3 cycles long (Section~\ref{multi_level_hypotheticals}), the Plan phase ultimately produces 1884 simulations (decision tree nodes) with 1728 unique decision tree paths.
Of the 1728 paths, there are 104 that satisfy the most \textit{:OptimalConditions} (5), and of those, there is 1 that most effectively optimizes the highest number of properties (1).

Lastly, the Execute phase enacts the decision and sets \textit{:Heater's} state to ``2'', \textit{:FloorHeating's} to ``0'', and \textit{:Dehumidifier's} to ``0''.
Since cycle durations are set to 900s (15 minutes), the RDT waits for the remainder of the time (\(900s - current\ cycle\ execution\ time\)) before starting the following cycle.


\section{Validation}
\label{validation}
To corroborate the contributions stated in Section~\ref{contributions}, we subjected our proof of concept to a series of empirical validations.
The following section splits the validation into two main parts, each involving a different TT and purpose: (\textit{i}) a virtual incubator setup and (\textit{ii}) a virtual office room M370 environment.
Due to the sheer number of possible RDT functionality configurations (Section~\ref{m370_example_demonstration}), our experiments only include selected graphs related to temperature and total energy prices incurred.
The remaining graphs related to different functionality combinations, as well as those illustrating humidity and energy consumption, can be found together with all of the experimental data in our public repository.\footnote{\url{https://github.com/ivanspajic/ruleless-digital-twins}}

\subsection{Incubator}
\begin{figure}[tbph]
  \begin{minipage}[b]{0.45\linewidth}
    \includegraphics[width=1\linewidth]{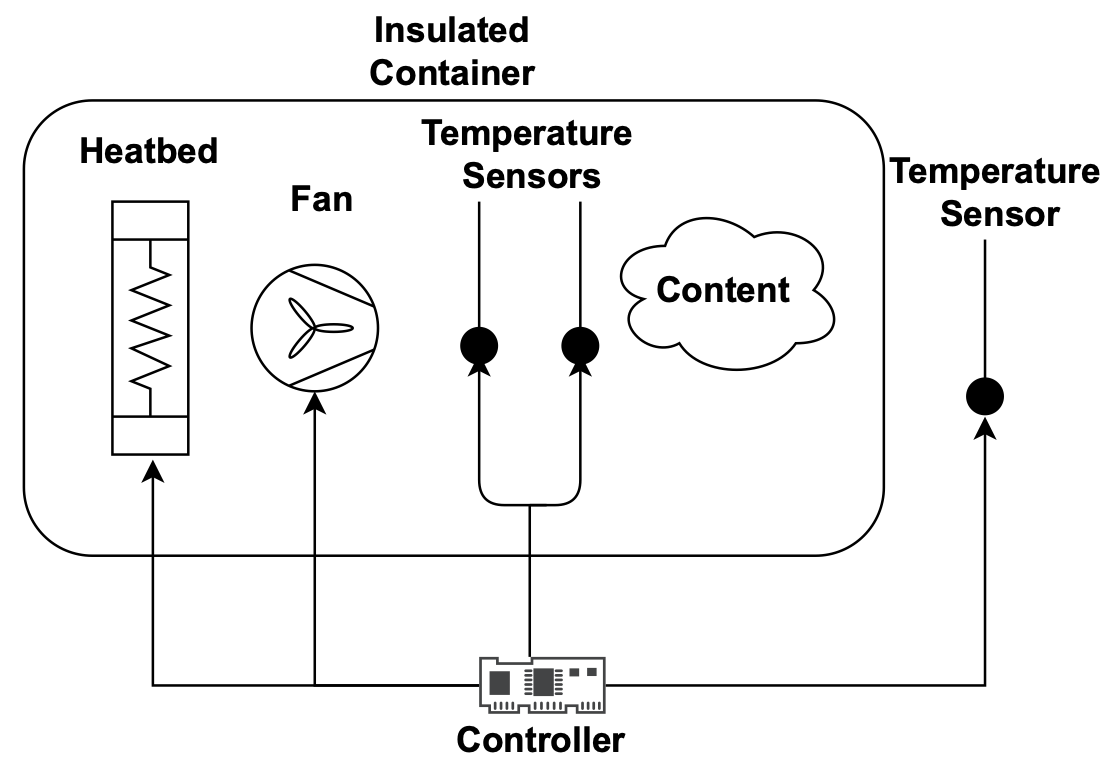}  
    \caption{Schematic overview of the incubator, reproduced from \cite{feng_et_al_2022}.}
    \label{fig:incubator_feng}
  \end{minipage}\quad
  \begin{minipage}[b]{0.5\linewidth}
    \includegraphics[width=1\linewidth]{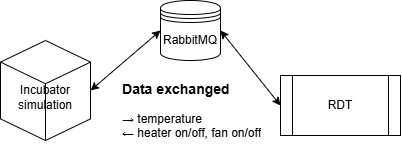}
    \caption{Communication between RDT and Incubator simulator.}
    \label{fig:rabbitmq}
  \end{minipage}
\end{figure}

We use a well-known digital twin example, Feng \emph{et al.}'s Incubator system \cite{feng_et_al_2022}, to illustrate the construction process of a model.
That our RDT with this model is in fact suitable to control a cyber-physical system in reactive mode with a tight cycle time, is an incidental byproduct.
It is not our goal here to compare the performance or the controlling strategy of the RDT against a reference controller.
Instead, it is sufficient that the RDT maintains the necessary optimal condition for the system purely from declarative specifications.

The incubator is designed to control a physical box consisting of a heater (to control temperature), temperature sensors (with the inside ones' averages used to drive the controller, while the last senses the static ambient temperature), and a fan that distributes heat evenly through the box (Figure~\ref{fig:incubator_feng}).
The goal of the controller is to provide optimal conditions for a fermentation-driven biological process requiring an ideal ambient temperature.
The box also has a lid (not shown) not controlled by the RDT but that rather models human interference with the system: if its state is ``open'', heat dissipates much more quickly.

While Feng et al. have developed and published the necessary materials to build this hardware controller, they also provided a comprehensive simulation environment for the fermentation box.
It contains the required machinery to compute realistic simulated sensor values given a configuration of settings consisting of current various temperatures, current temperature of the heating element, whether it is on or off, whether the fan is running, and whether the lid is open.
The environment comes with a simple MQTT setup to publish sensor values and receive new settings for the actuators, including additional useful machinery for logging and evaluating performance.
It is thus a drop-in replacement for the physical system which may, in fact, be used together with the hardware controller to control a physical experiment remotely.

We are using the simulation environment here which consists of a container-based deployment of their Python-based controller, a Python-based simulator for the physical components, and a message bus tying the components together (Listing~\ref{lst:plant}).
From their containerized default setup, we disconnect \textit{their} default Python controller (which reads sensor values and writes actuation commands) and connect \textit{our} RDT in its place.
We keep \textit{their} simulation component that runs in parallel and periodically computes and publishes the sensor-values.
This example is thus a manageable representative without depending on hardware (a digital TT, so to speak).
Although the behavior of the system is, of course, still computed virtually, we consider this simulation is a trustworthy proxy for the real system and not under our control.
Clearly we could explore scenarios such as opening the lid for which the DT does not have a sensor,
though we might be able to draw conclusion from the performance of the heater.
The communication through the broker is illustrated in Figure~\ref{fig:rabbitmq}.

Consequently, we now need the ontological model for this system and design the interface between it and the RDT.
The ideal operating range of the incubation process is represented by a single \textit{:OptimalCondition}.

Each MAPE-K round with a configurable interval (e.g., every 5 seconds), the RDT will need to inspect the current sensor readings and send actuations via the broker.
This requires a single, small, broker-specific .NET class which, through a simple configuration file specifying the address of the broker and a mapping between sensor/actuator-events and their respective topics and message format, needs to be connected to the respective components in the model.

\begin{figure}[tbph]
  \begin{minipage}[b]{0.56\linewidth}
\begin{lstlisting}[label={lst:incubator}, caption={Incubator-model fragment using the Python bindings that generate the RDF}, numbers=none, language=Python, basicstyle=\tiny\ttfamily]
MINE = Namespace("http://www.semanticweb.org/vs/ontologies/2025/12/incubator#")

temperature = ObservableProperty(g, MINE["T"], None)

heaterChange = Change(g, MINE["HeaterChange"], temperature,
                         increase=True)
heater = Actuator(g, MINE["HeaterActuator"], heaterChange,
                     actuatorName="in_heater_state",
                     actuatorType=XSD.integer,
                     actuatorStates=[0,1])
fan = Actuator(g, MINE["FanActuator"], None, # Pin fan to "on"
                  actuatorType=XSD.integer, actuatorStates=[1])

tempSensor = Sensor(g, MINE["TempSensor"], [temperature])

minTemp = Property(g, MINE["TemperatureLowerLimit"], 30)
maxTemp = Property(g, MINE["TemperatureUpperLimit"], 35)

oc_rtemp = OptimalConditionDouble(g, MINE["oc_temp"],
                              temperature, minTemp, maxTemp)

room = Platform(g, MINE["IncubatorTest"], False, [heaterActuator,
                   tempSensor, fan], implements=[oc_rtemp])
fmu = FMU(g, MINE["Incubator_FMU"], "Source/au_incubator.fmu", 3)
room.addFMU(g, fmu)
\end{lstlisting}\quad
  \end{minipage}
  \begin{minipage}[b]{0.41\linewidth}
    \vspace*{1cm} 
    \begin{lstlisting}[label={lst:plant},caption={Active components of incubator ({\small\texttt{start\_all\_services.py}})}, numbers=none, language=Python, basicstyle=\tiny\ttfamily]
# From https://github.com/INTO-CPS-Association/example_digital-twin_incubator:
      
 start_docker_rabbitmq()
 start_docker_influxdb()

 start_as_daemon(start_incubator_realtime_mockup)
 start_as_daemon(start_low_level_driver_mockup)
 start_as_daemon(start_influx_data_recorder)

# Simulated incubator:
start_as_daemon(start_plant_simulator)
# Disconnected controller:
# start_as_daemon(start_controller_physical)
    \end{lstlisting}
  \end{minipage}
\end{figure}

For our RDT to predict the different outcomes (e.g., the effect of the heater being on/off or the lid being open/closed), we need to provide a corresponding FMU.
We have opted to encode Feng et al.'s four-parameter equational system that their (disconnected) Python controller uses as an OpenModelica FMU.

Instead of writing a Turtle file directly, we use a more concise and readable Python object-relational mapping that will produce it, thus cutting down file size from 84~lines for the generated Turtle file to 40~lines in the Python bindings (Listing~\ref{lst:incubator}).
\vsnote{Could also have an object diagram here, would take about the same space. But I want to show what the user \textbf{actually} needs to write, not some abstraction.}

First, we declare our namespace and then create the desired observables, sensors, and actuators.
The heater has states off and on (modeled through values 0 and 1) and is annotated as increasing the temperature in the box.

The fan is intended to run permanently but needs to be switched on in the beginning, so we model this using an actuator with a single action.
We declare the optimal temperature range as 30--35°C, tie all elements together in the \textit{sosa:Platform} (root) node of the model, and attach the FMU.
Note that inputs/outputs of the FMU are matched implicitly by name unless overridden, such as \texttt{T} and \texttt{in\_heater\_state} being present in both the instance model and FMU.
Not shown is the property \texttt{power\_out\_box} which is synthesized from the FMU and reports the predicted power consumption for the heater.
The lid is not an element of the model, since it is manipulated outside of the control of the controller, does not have a sensor, and only affects the behavior of the simulated systems (not the simulation in the DT):
it changes a parameter in the simulated system, making heat dissipate faster, and can be manipulated interactively from the command line.

\begin{figure}[tbph]
\begin{lstlisting}[numbers=none, basicstyle=\tiny\ttfamily]
Logic.Mapek.IMapekPlan: Information: Total simulation time (seconds): 0.909
Logic.Mapek.IMapekPlan: Information: Generated a total of 8 simulation paths.
Logic.Mapek.IMapekPlan: Information: Chosen optimal path, Actuation actions:
Interval 1:
Actuator: http://www.semanticweb.org/vs/ontologies/2025/12/incubator#FanActuator, Actuator state: 1.
Actuator: http://www.semanticweb.org/vs/ontologies/2025/12/incubator#HeaterActuator, Actuator state: 1.
...

Logic.Mapek.IMapekExecute: Information: Starting the Execute phase.
Logic.Mapek.IMapekExecute: Information: Actuating actuator http://www.semanticweb.org/vs/ontologies/2025/12/incubator#FanActuator with state 1.
Logic.Mapek.IMapekExecute: Information: Actuating actuator http://www.semanticweb.org/vs/ontologies/2025/12/incubator#HeaterActuator with state 1.
Logic.Mapek.IMapekExecute: Information: Expected Property values:
http://www.semanticweb.org/ivans/ontologies/2025/instance-model-1#MapekCycle: 15
http://www.semanticweb.org/vs/ontologies/2025/12/incubator#TempMeasure: 33.36249289894771
http://www.semanticweb.org/vs/ontologies/2025/12/incubator#T: 33.51256725018559
...
Logic.Mapek.IMapekManager: Information: Sleeping 5000 ms until next MAPE-K cycle.
\end{lstlisting}
\caption{Output of RDT when controlling the incubator.}
\label{lst:inc_mapek}
\end{figure}

Figure~\ref{lst:inc_mapek} shows the partial output of a MAPE-K round taken by the RDT.
Given that there are only two possible actuations in each state (heater on or off; as the fan is fixed) and that we simulate a tree of depth 3, we obtain \isnote*{Then we're definitely not reactive!}{8 paths of lengths 3 each},
from which the RDT picks the optimal and enacts the actuations from its first decision.
Of note is that while the system is within the optimal condition (\texttt{TempMeasure} is the input temperature, \texttt{T} is the predicted temperature given the chosen actions by the FMU),
many different plans may perform equally.
In a configuration that the RDT should look ahead for three five-second cycles, either choice of setting the heater to on or off will keep the system within its optimal condition.
In the absence of any other constraints, the RDT will make a non-deterministic choice.
Only when the simulation indicates that a particular choice will score negatively wrt.\ the \textit{:OptimalCondition}, will the RDT prefer a particular choice, if any.
Executing the complete round with simulations through the FMU and picking a candidate decision took less than a second on our desktop computer.
\vsnote*{Could also/additionally go into the conclusion.}{%
It is clear that there is a lower performance limit to deploy an RDT with its combined .NET and Java architecture which might not be ideal for resource-constrained targets.
}%

In summary, this case study shows the necessary steps for the end-user to obtain a working digital twin:
\begin{itemize}
\item provide the instance model with the static components;
\item add \textit{:OptimalConditions} and optional optimization target properties;
\item annotate optional effects in the model (heater increases temperature and energy consumption);
\item provide the the FMU; 
\item configure adapters to inject sensor values and enact actuations;
\item choose RDT configuration parameters (e.g., prediction horizon and interval length).
\end{itemize}
It is obvious that complementary expertise is required:
whereas we feel that modeling the first three steps (either through a future IDE or suitable bindings like in Python) might be possible for non-expert end-users, the development of FMUs in particular requires both domain and programming knowledge.
For the adapters, we foresee that we can eventually provide a library of common connectors to, e.g., REST APIs, MQTT, or even GPIO,
that only need to be instantiated with suitable parameters, closely resembling how smart home owners already configure their systems.



\subsection{Virtual Office Room M370}
\label{validation_m370}
The virtual M370 office room environment is implemented as a rudimentary wrapper around the simulation model representing an HA version of its real-world counterpart.
All relevant user-provided sensor and actuator connection adapters in the RDT couple to the virtual environment to facilitate observing and enacting changes on it, respectively.

Although we utilize the same FMU to run both the virtual environment and the RDT's Plan phase simulations for M370, the wrapper implementation includes an element of (seeded) randomness to account for any aleatoric uncertainty.
The focus of these experiments is neither on cyber-physical systems nor on the accuracy of our simulation models but instead on the demonstration of the advantages and disadvantages of the RDT concept as a method of decision-making and system control.

\subsubsection{Experimental Setup}
\label{validation_m370_experimental_setup}
\begin{lstlisting}[label={lst:validation_bang-bang_rules}, caption={A C\# implementation of the rules governing the bang-bang controller's decision-making.}]
public class BangBangPlanner
...
if (officeTemperature <= MinimumOfficeTemperature) {
    heater.State = 1;
    floorHeating.State = 1;
} else {
    heater.State = 0;
    floorHeating.State = 0;
}

if (officeHumidity >= MaximumOfficeHumidity) {
    dehumidifier.State = 1;
} else {
    dehumidifier.State = 0;
}
\end{lstlisting}

To compare our results against an experimental control, we developed a rule-based decision model in the form of a classic bang-bang controller.
It uses M370's \textit{:RoomTemperature} and \textit{:RoomHumidity} properties to perform relevant actuations in a simple ON/OFF fashion, entirely governed by its \texttt{IF-THEN} rules (Listing~\ref{lst:validation_bang-bang_rules}).

The goal of both the bang-bang controller and the RDT was to keep these properties within their respective \textit{:OptimalConditions}, namely \textit{:OfficeTemperatureOptimalCondition} and \textit{:OfficeHumidityOptimalCondition} (introduced in Section~\ref{optimal_conditions}), against a cold and humid virtual environment.
To contextualize our experiments accordingly (with an emphasis on MPC), we included the snapshot of Nord Pool's spot energy prices (Section~\ref{cycle_durations}) as part of our virtual TT and annotated its instance model with a minimization on the total price per energy consumed.
By doing so, we forced the RDT to conduct its decision-making under dynamic (continuously changing) conditions with limited predictability (bounded prediction horizon) \cite{camacho_model_2007, nikolaou_model_2001}.
The minimization on an ever-changing property also appropriately highlights the contrast between enabling optimizations in the RDT versus doing so in the rule-based control (Section~\ref{problem_domain}).

All experiments related to temperature, humidity, energy consumption, and price per energy consumed\footnote{Note that graphs illustrating total energy consumptions and total energy costs start at 0 on the first (0th) cycle due to these properties depending on actuations in the Execute (last) phase of a MAPE cycle. Such values are thus first observed in the following cycle.} were conducted over 30 MAPE-K cycles with varying lengths of prediction horizons and degrees of property value deviation tolerance, as well as with and without reactive decision-making heuristics.
Similarly, we tested CBR functionality by running the RDT over 250 MAPE-K cycles, varying prediction horizon lengths and property value deviation tolerances, and starting each run with a clean case database.
Lastly, we gathered and analyzed performance data in terms of MAPE-K cycle durations for all of the experiments run, ultimately highlighting some of the concept's advantages and disadvantages.

\subsubsection{Multi-Level Hypotheticals Results}
\label{validation_m370_core_results}
\begin{figure}
    \begin{subfigure}{0.5\textwidth}
        \includegraphics[width=1\linewidth]{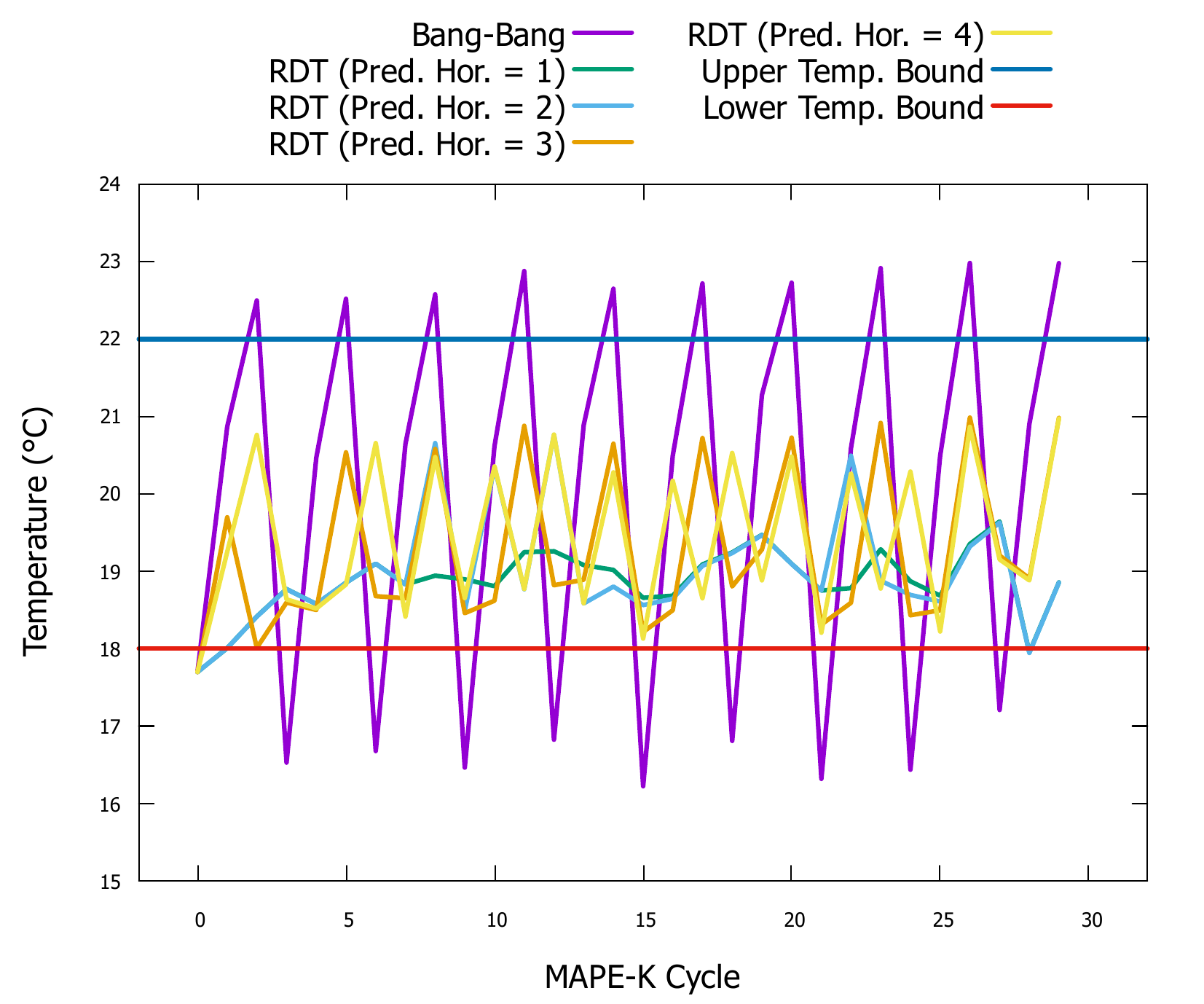}
    \end{subfigure}
    \begin{subfigure}{0.5\textwidth}
        \includegraphics[width=1\linewidth]{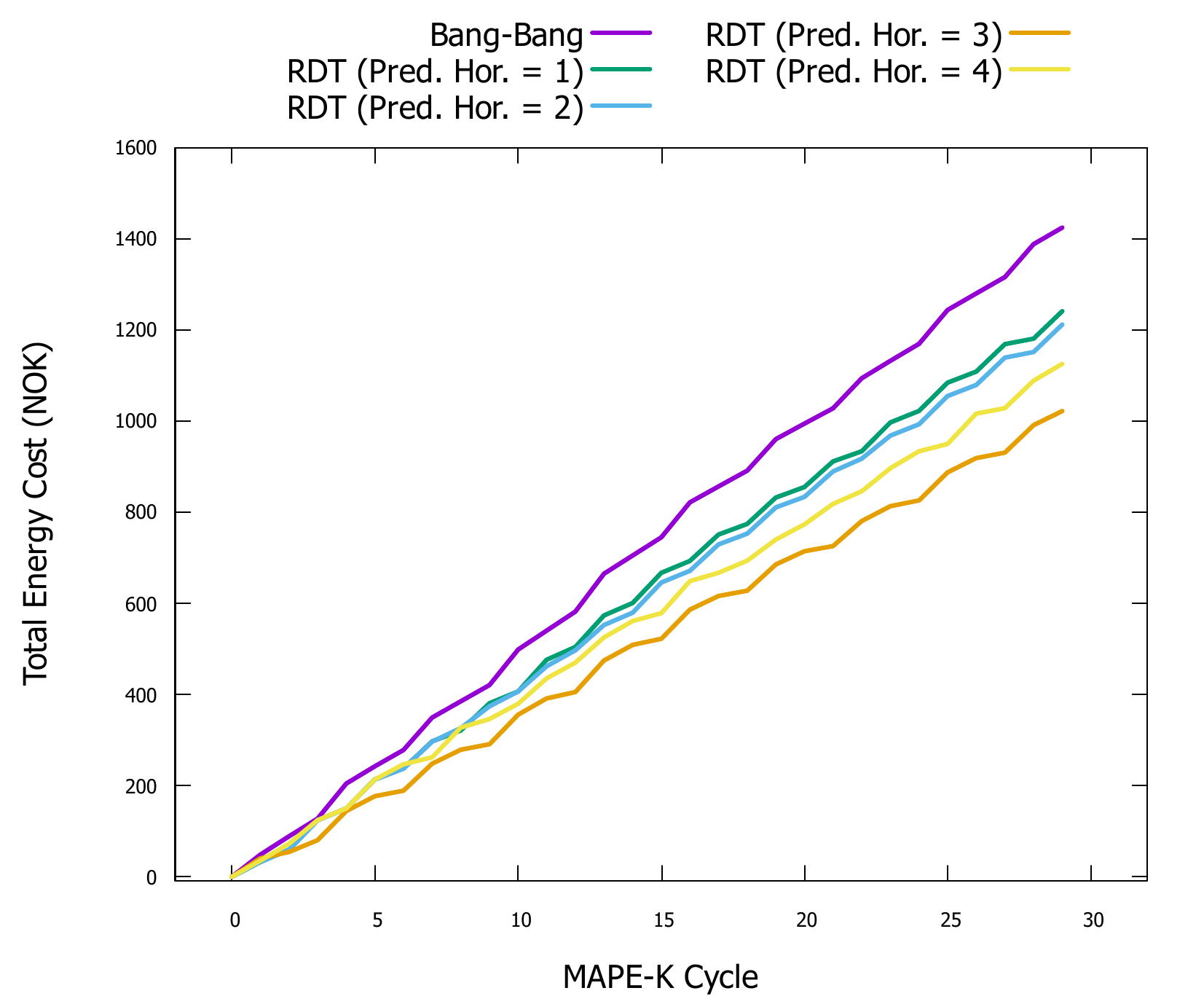}
    \end{subfigure}
    \caption{Room temperatures and total energy costs of bang-bang, RDT-pred-1, RDT-pred-2, RDT-pred-3, and RDT-pred-4 over 30 MAPE-K cycles.}
    \label{fig:bang-bang_rdt_ph1_ph2_ph3_ph4}
\end{figure}

Figure \ref{fig:bang-bang_rdt_ph1_ph2_ph3_ph4} illustrates room temperatures kept and total energy costs incurred by the bang-bang controller (bang-bang) and RDTs with different prediction horizon lengths: 1 cycle (RDT-pred-1), 2 cycles (RDT-pred-2), 3 cycles (RDT-pred-3), and 4 cycles (RDT-pred-4).
All RDTs used a property value deviation tolerance of 2 with none of them using reactive decision-making heuristics or CBR.

Notably, the RDTs kept a more stable temperature (almost entirely within bounds) with RDT-pred-1 keeping it most stable.
Since the total price incurred directly correlates with the total energy consumed, all RDTs also generally kept their respective temperatures closer to the lower bound, resulting in bang-bang being most expensive (1423.58 NOK).
Moreover, it is interesting to observe how increasingly longer prediction horizons resulted in RDTs automatically inferring increasingly more fluctuant strategies, subsequently incurring lower costs.

Most interestingly, however, even though we expected to see RDT-pred-4 incurring the lowest total energy cost, due to having the longest prediction horizon, RDT-pred-3 was actually the cheapest controller over the full 30 cycles (1021.32 NOK).
A possible explanation is that although running with a prediction horizon of 4 allows a potentially greater total cost minimization than running with a prediction horizon of 3, provided they share a starting point (same cycle), it is entirely possible that running with a prediction horizon of 3 ultimately aligns more closely to the theoretical 30-cycle optimal once their starting points are \isnote*{Does this need an example, or is it understandable enough? I can put things in terms of 5 cycles if need be.}{different}.
This implies that RDT-pred-4 could have had better optimizations had it known which starting points (cycles) were best for a prediction horizon of 4 over the whole 30 cycles, thus also requiring the option of not executing all of its buffered decisions but choosing to simulate early instead.
Given that it can only look ahead for 4 cycles at a time (and accept all of its buffered decisions if they match their conditions), it only produced decisions optimizing costs over those 4, regardless of its starting point.

\subsubsection{Reactive Decision-Making Heuristics Results}
\label{validation_m370_reactive_results}
\begin{figure}
    \begin{subfigure}{0.5\textwidth}
        \includegraphics[width=1\linewidth]{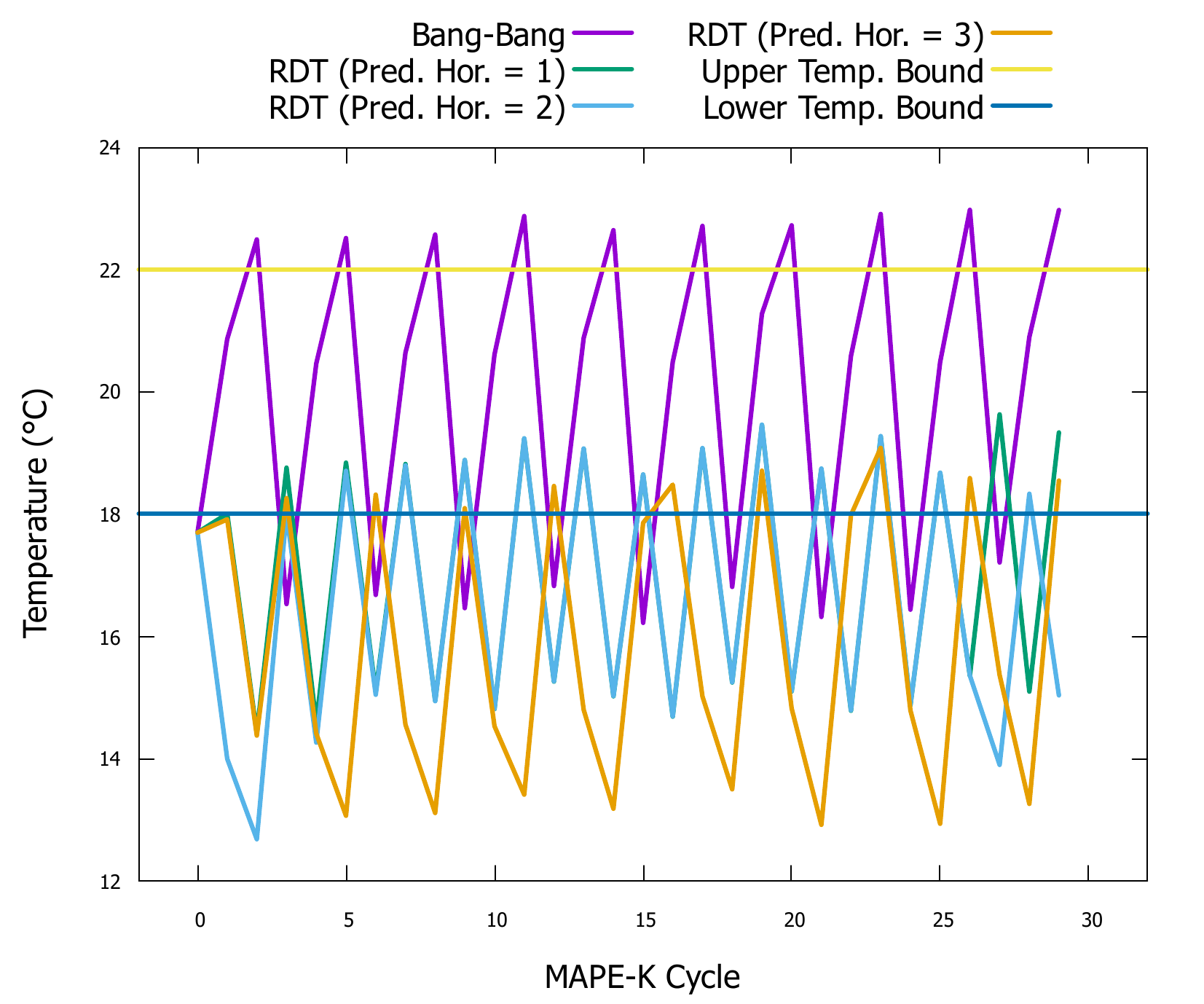}
    \end{subfigure}
    \begin{subfigure}{0.5\textwidth}
        \includegraphics[width=1\linewidth]{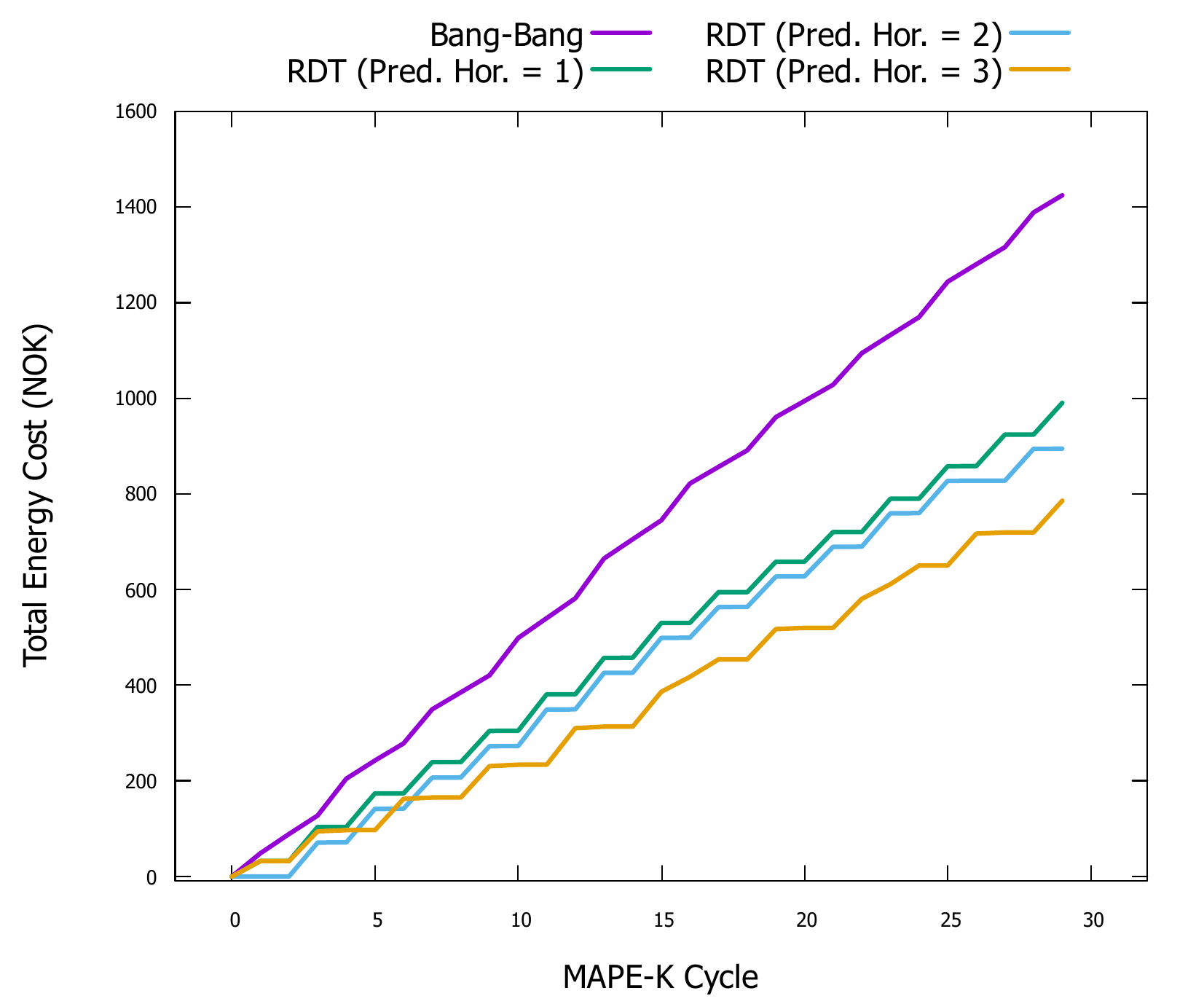}
    \end{subfigure}
    \caption{Room temperatures and total energy prices of bang-bang, RDT-react-1 and RDT-react-2 over 30 MAPE-K cycles.}
    \label{fig:bang-bang_rdt_react_ph1_ph2}
\end{figure}

Figure \ref{fig:bang-bang_rdt_react_ph1_ph2} compares bang-bang with three RDTs: the first one using a prediction horizon of 1 (RDT-react-1) the second that of 2 (RDT-react-2), and the third that of 3 (RDT-react-3), with all three using reactive decision-making heuristics and a property value deviation tolerance of 2.

Although all RDTs incurred substantially lower costs than bang-bang, it is evident they kept M370's temperature well below the lower bound for most of the 30-cycle duration.
To that end, the constantly cooling virtual environment perfectly highlights the drawbacks of purely reactive decision-making heuristics.
Because all RDTs only inferred possible decisions when \textit{:OptimalConditions} were violated, no actions were taken when the temperature was (barely) within bounds, thus allowing it to plummet once again.

The same is evident with an increased prediction horizon, hence demonstrating a matching graph: regardless of being able to look ahead for multiple cycles, as soon as RDT-react-2 and RDT-react-3 predict M370's temperature to be within \textit{:OptimalCondition} bounds, they will infer no more possible decisions for any subsequent ones.

\subsubsection{Case-Based Reasoning Results}
\label{validation_m370_case-based_reasoning_results}
\begin{figure}
    \centering
    \includegraphics[width=0.75\linewidth]{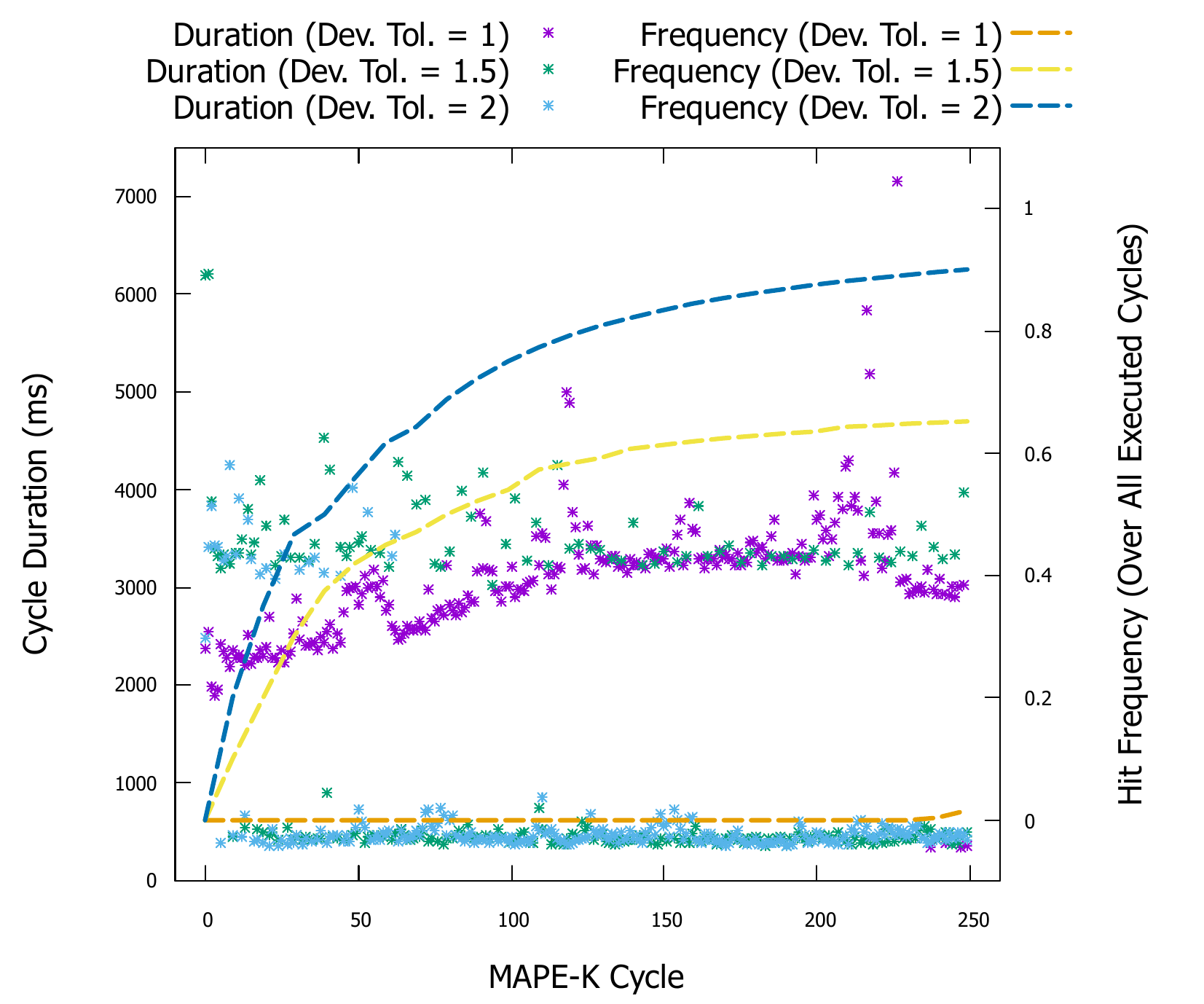}
    \caption{Cycle durations and case hit frequencies of RDT-dev-1, RDT-dev-1.5, and RDT-dev-2 over 250 MAPE-K cycles. The longer duration grouping indicates no matches, while the shorter one indicates matches.}
    \label{fig:case_hit_frequencies}
\end{figure}

Figure \ref{fig:case_hit_frequencies} contains two parts: firstly, it shows cycle durations of three separate RDT runs with CBR functionality and a prediction horizon of 1, whereby the first one used a property value deviation tolerance of 1 (RDT-dev-1), the second that of 1.5 (RDT-dev-1.5), and the third that of 2 (RDT-dev-2).
Secondly, in steps of 10 for the duration of the run, it illustrates the frequency of case hits over the number of all (up to that point) previously executed cycles.

This demonstrates two closely related concepts: cycle durations reveal not just that a case had been matched on but also the direct performance gain as a result thereof.
Moreover, the differences in property value deviation tolerance emphasize its impact on the number of MAPE-K cycles an RDT must complete to (increasingly) reap the benefits of its CBR.

RDT-dev-1 had only 4 hits after the 238th cycle, RDT-dev-1.5 had scattered hits (relatively evenly distributed), and RDT-dev-2 had its last miss on the 63rd cycle, with all configurations showing variance in both hit frequencies and their respective rates of change.
Because larger deviation tolerances allow saved cases to cover more of a given TT property's value domain, it is entirely expected that they facilitate more frequent hits much faster.
Lastly, all configurations continuously maintained their respective room temperatures within \textit{:OptimalCondition} bounds, with RDT-dev-1 and RDT-dev-1.5 demonstrating 1-2 short-lasting, shallow dips below the lower bound.

It should be noted that our implementation of CBR specifically only deals with relevant properties.
Our cases thus save, and match on, a subset of a TT's properties for two reasons: firstly, certain cumulative properties, such as total energy consumption or total energy cost, can only change monotonically.
Including them as part of a case would obviously make matching at a later cycle difficult, even if all other observable properties, such as temperature or humidity, proved near identical.
Secondly, because property value observations happen at the beginning of a cycle, properties that rely on executions of actuations, such as the current (non-total) energy consumption, as well as anything derived from it, can only be observed in the cycle that follows.
Moreover, because case reuse aims to skip the Analyze and Plan phases by checking for a possible match (after the Monitor phase), any hits would require that the case has a similar-enough energy consumption value to that resulting from the Execution phase of the cycle before.
Therefore, to avoid implicitly binding a case to its preceding cycle, such (and any closely related) properties are ultimately excluded (discussed further in Section~\ref{future_work}).

\subsubsection{RDT Performance}
\label{rdt_performance}

To test the performance of our proof-of-concept RDT, we sampled cycle duration data (elapsed wall-clock time per cycle) for various configurations over 30-cycle runs using a Windows 11 machine equipped with a 12th generation Intel Core i5-1235U.
These are only given for completeness as our implementation is not tuned for performance: we neither make use of multi-threading nor optimize the frequent and expensive on-disk instance model exchange with the inference engine,
which currently accounts for more than 90\% of reactive heuristics execution time.

\begin{table}[tbph]
\centering
\caption{Average cycle durations (in seconds) over 30 MAPE-K cycle runs.}
\label{tab:rdt_performance}
    \begin{tabular}{ccccccccc}
    \hline
    & \multicolumn{8}{c}{\textbf{Prediction Horizon (Dev. Tol. 1 \& 2)}} \\
    & \multicolumn{2}{c}{\textbf{1}} & \multicolumn{2}{c}{\textbf{2}} & \multicolumn{2}{c}{\textbf{3}} & \multicolumn{2}{c}{\textbf{4}} \\
    \hline
    \textbf{Core}  & 2.82 & 3.52 & 2.31 & 1.70 & 10.95 & 4.80 & 62.69 & 48.49 \\
    \textbf{React.} & 1.95 & 2.00 & 12.63 & 3.46 & 99.12 & 26.27 & * & * \\
    \hline
    \end{tabular}
\end{table}

Table \ref{tab:rdt_performance} shows a general trend of increasing MAPE-K cycle durations consistent with longer prediction horizons.
Both core and reactive functionality exhibited shorter cycle durations with prediction horizons of 2 versus those of 1, with the exception of the prediction horizon of 1.
Although there is a general trend of higher deviation tolerance having shorter average cycle durations, this is to be expected.
Naturally, since higher deviation tolerance permits for easier property value matching, any buffered decisions (from longer prediction horizons) will have higher chances of being used, subsequently allowing the RDT to skip its Analyze and Plan (slowest) phases.
Lastly, reactive decision-making heuristics proved generally slower than the RDT's core functionality.\footnote{For performance reasons, due to wide trees, reactive decision-making heuristics were tested with a maximum prediction horizon of 3.}
Although we expected to see performance improvements due to the heuristics' tree-pruning strategy, we can confidently attribute this slowness to its more frequent uses of the (relatively slow) inference engine through the .NET profiler.

The performance data demonstrates the existence of a kind of trade-off between the level of optimization an RDT may achieve (through its prediction horizon length) and the types of domains it is applicable to.
In some domains, such as HA, fitting cycles (of up to 3--4 minutes) into 15-minute intervals for climate control is entirely appropriate.
Conversely, other domains may require cycles to fit within near real-time intervals (e.g., \( <5 \) seconds), forcing the RDT to only achieve limited optimizations.
Although RDTs could certainly be optimized for performance, this ultimately imposes a prerequisite on end-users to independently verify their feasibility in the context of particular requirements and application domains.

\subsubsection{Threats to Validity}
\label{validation_m370_threats_to_validity}
The following sections outline various threats to the validity of our experiments and the conclusions drawn therefrom \cite{shull_selecting_2008}.

\paragraph{Internal Threats to Validity}
Despite the declarative nature of the RDT as a software product, we have not proven its implementation as correct.
Although a number of test cases check expected behavior, in general, ground truth exists only on a per-model basis, which includes their FMUs operating as black boxes on floating point values.
Due to non-determinism in our codebase, e.g.\ when choosing among several equally ranked---but not equivalent---decisions,
it is sometimes very hard to verify or even interpret the behavior of a simulation after the fact.
In principle, programming mistakes are hard to discover in such a system.


\paragraph{External Threats to Validity}
One potential threat to external validity is that of having used a virtual environment as a TT for our experimental evaluations.
Even though we introduced randomness to the environment, its underlying mechanism still relies on the very same simulation model used by our RDT, therefore naturally providing it with a high degree of prediction accuracy.
Because an RDT, like any DT, relies on simulations to assess hypothetical scenarios, we implicitly assume the availability of sufficiently accurate models that enable and provide such functionality.
It is not our intention to construct or evaluate simulation models according to their high fidelity but to present RDTs that use such models as viable concepts of \textit{ruleless} decision-making.
This same assumption exists in the field of MPC, where accurate system control inherently relies on accurate prediction models \cite{camacho_model_2007}.
Even with an accurate model (and slight randomness), our experiments show that RDTs do not always match on their previously buffered decisions, subsequently still demonstrating the concept's adaptability and resilience, even when facing inaccurate predictions.
To that end, by reasonably assuming accurate simulation models, we strictly emphasize RDTs as feasible \textit{ruleless controllers}, regardless of their observed data source, and not necessarily as DTs tailored to particular instances of physical office rooms.

Another potential threat lies in the applicability of RDTs to domains other than HA.
Although our experiments were only conducted in the context of HA, there is nothing domain-specific in the ontology, inference rules, or FMUs as a simulation framework.
As demonstrated through the incubator example, all domain-specific elements of an RDT are confined to the instance model reflecting its TT.
Since there is nothing inherently domain-specific to an RDT's decision-making method, users are free to employ the concept as long as its performance adheres to their domain-specific constraints (discussed in Section~\ref{rdt_performance}).

\paragraph{Construct Threats to Validity}
A possible threat to construct validity might exist in the lack of tested configurations and choice of run durations.
Despite a finite number of experiments, we demonstrate how even core RDT functionality (Section~\ref{contributions}) maintains conditions with respect to user specifications without using any rule-based decision-making models (shown in Section~\ref{validation_m370_core_results}).
Additionally, our 30-MAPE-cycle runs represent 7.5 hours in 15-minute intervals, which is large enough to be non-negligible in illustrating the continuous satisfaction of \textit{:OptimalConditions}.
Similarly, our 250-cycle CBR runs were sufficient in highlighting the impacts of different property value deviation tolerances on case hit frequencies and performance.


\section{Discussion}
\label{discussion}
Although RDTs rely on inference rules to expand the semantics of our ontological concepts (and enable reactive heuristics), their \textit{rulelessness} ultimately comes from users no longer needing to design and maintain rule-based models for dynamic TTs.
Our concept therefore effectively streamlines the domain of user-provided artifacts by reducing complex, decision-making rules to simple, declarative specifications.
Crucially, instead of describing \textit{undesirable} TT conditions (through the \textit{CONDITION} part of a rule), such specifications allow users to directly describe the \textit{desirable} ones.

Similarly to MPC utilizing declarative constraints and a prediction horizon, RDTs make use of \textit{:OptimalConditions} and multi-level hypotheticals.
The process of finding an RDT's optimal decisions also closely resembles that of mixed-integer MPC: it begins with prediction models (simulations) facilitating the construction of a tree of finite possibilities, whose root-to-leaf paths represent futures which can be ranked according to their \textit{:OptimalCondition} satisfactions.

Whereas mixed-integer MPC optimizations rely on algorithms guided by cost functions, such as branch-and-bound, to explore the decision tree state space, we do not yet implement such concrete optimization for large value domains.
The instance model annotations (\textit{:minimizes} and \textit{:maximizes}) can \vsnote*{new}{through their interaction with \texttt{:ValueIncrease/Decrease}-annotations limit the selection of actions}, but do not scale to e.g.\ integers or floating point values (discussed further in Section~\ref{future_work}).

Despite initially expecting simulations to be the primary source of longer cycle durations, performance experiments with reactive decision-making heuristics revealed this to actually be the inference engine.
Although the heuristics correctly pruned decision trees through their inference rules, any gains from reduced numbers of simulations were negated by employing inference at every level in the prediction horizon.
Moreover, experiments demonstrated that a purely reactive approach to TT control combined with property value minimizations was less than adequate.
Unlike an equally reactive bang-bang controller, a purely reactive RDT had the ability to restore property values to their \textit{:OptimalConditions} but lacked the possibility of preemptively maintaining them there.

To enable decision generation and reactive heuristics, we had to extend the SSN ontology.
Because we performed no alterations to any of its original triples, we demonstrated the possibility of extending an ontology to support additional reasoning through custom inference rules.
This enables users of RDTs to further expand the semantic knowledge base with their own ontologies to power an additional or altogether different inference process.
For example, certain actuator states might not be reachable from certain others, potentially requiring more specific ontological classes and/or inference rules to perform such checks.
Therefore, instead of having to write imperative reasoning algorithms as part of a codebase, users can simply utilize declarative rules to develop custom decision-yielding heuristics, although the current inference engine leaves much to be desired in the way of performance.

One of the biggest advantages of RDTs lies in their flexibility.
By utilizing the MAPE-K architecture, they leverage strengths from autonomic computing to dynamically adapt to changing TT conditions and user specifications.
There are, however, potential pitfalls: during especially prolonged simulation times (e.g., due to a longer prediction horizon), TT conditions may change drastically, particularly through unforeseen real-world events.
To that end, a more event-driven architecture allowing for interruptions would ensure the DT discards hypothetical decisions based on newly stale or obsolete TT conditions.

RDTs unfortunately also require \isnote{Following Merriam-Webster everywhere.}{premade} FMUs encompassing all observable properties and actuators in their TTs.
This could, however, be considered a trade-off in development time otherwise traditionally spent on imperative, rule-based decision-making models.
Since they are decoupled from user specifications, such as \textit{:OptimalConditions}, they arguably provide greater exchange potential compared to traditional rules.

Naturally, another prerequisite is that of determining the applicability of RDTs to a given problem domain with respect to their performance.
Given the inherent correlation between prediction horizon lengths and MAPE-K cycle durations, users are tasked with understanding whether the highest levels of optimizations possible are adequate under the constraints of their domains.

Note that our system is not an alternative to machine learning.
On the contrary, RDTs may be used as a precondition: in the early stages of deployment and after reconfiguration, RDTs provide the desired and transparent behavior and hence the ``ground truth'' necessary to train a machine learning model, which could subsequently be used to shortcut decision-making, similarly to CBR.
It is not immediately obvious if the simulations in themselves are useful enough to create training data for machine learning due to their (perceived) high computational cost.



\section{Related Work}
\label{related_work}
Kamburjan et al. \cite{kamburjan_et_al_2021} use ontologies to map program state onto semantic knowledge graphs.
Although their focus is on debugging and program state validation, RDTs similarly combine observed property values with a static ontological instance model to create a dynamic one, subsequently affecting inferrable knowledge.
Their presented language, SMOL, supports integration between semantic assets, such as our ontological instance models, and Functional Mock-up Objects (FMOs) acting as wrappers around FMUs.
Users can thus employ SMOL to construct DTs and their decision-making by directly integrating their semantic and simulation models.
RDTs, on the other hand, allow users to construct DTs through simple semantic instance models, whereby the mechanics of decision-making are automatically handled.

Compagnucci et al. \cite{compagnucci_et_al_2023} use the SSN ontology as a semantic foundation for developing structural TT models for their model-driven DT implementation.
Although their paper discusses the implications of inference strategies, they do not employ them in their approach.
While their prototypical DT implementation does require a human-in-the-loop to facilitate manual communication between the DT and the TT, they effectively automate decision-making with finite state machines.

The utilization of ontologies also enabled the onset of the rather controversially named \textit{cognitive digital twins} (CDTs) \cite{zheng_emergence_2022}.
Although RDTs use a semantic knowledge base to enrich their instance models and enable decision-making heuristics, some of the wide-spanning definitions of CDTs also cover system evolution \cite{abburu_cognitwin_2020}, which is not the focus of our work.
Despite the autonomous decision-making ability of RDTs, we refrain from using the term ``cognition'' as we do not see evidence of functionality beyond what we already present.

Korth et al. \cite{korth_et_al_2018} demonstrate an event-driven architecture combining a logistics system DT with simulation capabilities in the context of a DSS with an human-in-the-loop.
Unlike the RDTs' MAPE-K-based approach, their architecture enables for more dynamic system responses due to its event-driven nature.
This allows sudden and drastic TT condition changes to potentially interrupt ongoing DT executions to trigger new simulations.
Crucially, however, our RDTs follow their proposed list of five necessary DSS criteria: (\textit{i}) knowledge of current system states, (\textit{ii}) knowledge of future system states, (\textit{iii}) functions identifying current and future system disruptions, (\textit{iv}) functions generating decision scenarios, and (\textit{v}) functions evaluating such scenarios.

Päßler et al. \cite{paesler_et_al_2022} propose a formal model of an IoT-based architecture for smart home solutions.
They automate decision-making by defining rule-based models for different kinds of actuator controllers and show how a combination of these yields improved performance with respect to requirements.
Similarly to RDTs, their implementation utilizes a formal, ontological model for describing system components and user specifications, as well as a MAPE-K loop that uses the model as its knowledge base.
However, since their implementation and experiments were realized exclusively in Maude, they leave little room for additional, user-provided specifications or controllers.

Listl et al. \cite{listl_et_al_2021} present the idea of combining components shared by RDTs to enable DTs to be used as DSSs.
Firstly, they describe data provisioners and mappers for the purposes of data extraction and modeling.
Listl et al. also incorporate semantic knowledge bases and simulation models as key elements in DT decision-making.
However, since they primarily intend to utilize semantic knowledge bases for the purposes of standardizing data exchange and relating TT components to one another, they ultimately do not employ logical inference as part of their process.
Lastly, Listl et al. also outline multiple types of analytics covering current scenario conditions and predicted decision outcomes.

Bolender et al. \cite{bolender_et_al_2021} apply DTs with CBR to managing unforeseen scenarios with injection molding machinery.
They use model-driven methods to design languages for defining decision rules and case similarity.
Although their framework supports the added functionality of adapting existing cases to match newly-encountered conditions, it does not construct them nor their rules automatically.
Michael et al. \cite{michael_et_al_2024} utilize CBR in a similar manner to enhance decision-making explainability in their model-driven DT architecture.
Much like our RDTs, their custom framework is also built on the principles of the MAPE-K loop.

Yigal et al. \cite{yigal_et_al_2024} propose an approach of managing multiple sources of HA rules through a three-level feedback loop architecture.
Their \textit{primary outer loop} determines what properties should be affected, their \textit{middle secondary loop} what devices should be actuated, and their \textit{tertiary innermost loop} what manner to actuate them in.
Each level contains a specific manager with knowledge pertinent to its own functionality.
As such, much like with RDTs, users are ultimately only required to provide high-level, declarative specifications with the system taking care of the rest.
However, instead of hypothetical scenario exploration, the authors automate device-level decision-making in the so-called \textit{tertiary innermost loop} through the use of simple sensor-actuator feedback loops.

Walton et al. \cite{walton_et_al_2024} define a DT with real-time decision-making as a system exhibiting four capabilities; (\textit{i}) describing what has happened and what is currently happening, (\textit{ii}) predicting events that are likely to happen, (\textit{iii}) prescribing corrective actions, and (\textit{iv}) actualizing them.
Although RDTs are comparatively limited in predicting likely events, they cover the other three capabilities fully.
To power their decision-making process, Walton et al. employ a feedback mechanism similar to the MAPE-K by expanding Boyd's original Observe, Orient, Decide, and Act (OODA) \cite{richards_2020} loop to include a Simulate step.
They also consider decision lag mitigation (discussed in Section~\ref{future_work}), whereby a system attempts to account for state changes between initial data observations and the resulting final decisions.
Lastly, they state that autonomous DT decision recommendations can be facilitated by R\&S algorithms, similarly to how they are utilized by RDTs.

Kabashkin \cite{kabashkin_2025} presents an ontology-driven DT framework for explainable decision-making in aviation maintenance and operations.
He formally defines a DT as a 6-tuple consisting of a physical asset \( \mathcal{P} \), a set of input ontologies \( \mathcal{O}_{in}(t) \) at time \( t \), a set of output ontologies \( \mathcal{O}_{out}(t) \) at time \( t \), a set of semantic mappings and transformation rules \( \mathcal{M} \), reasoning mechanisms \( \mathcal{R} \), and a function \( \mathcal{U} \) for gathering observations and advancing DT state.
In terms of RDTs, \( \mathcal{P} \) represents the TT, \( \mathcal{O}_{in}(t) \) the pre-inferential knowledge base state (at time \( t \)), \( \mathcal{O}_{out}(t) \) the post-inferential knowledge base state (at time \( t \)), \( \mathcal{M} \) the logical inference rules, \( \mathcal{R} \) the inference engine, and \( \mathcal{U} \) the MAPE-K loop and our proof-of-concept framework.
Since he uses multiple ontologies in conjunction, Kabashkin mostly focuses on their semantic interoperability and consistency.
Although he discusses the implications of using logical inference\footnote{Kabashkin uses the term ``mathematical inference''.} for explainability and DT decision-making, his own work in the domain is limited to consistency checks and class subsumption inference with Protégé's \cite{musen_and_protege_2015} built-in reasoner.

Overall, it is evident that a plethora of past work focused on decision-making approaches through rule-based control, even in dynamic systems, such as manufacturing and warehouse logistics.
Our research thus attempts to advance beyond rule-based DT decision-making by introducing it to the well-established \textit{ruleless} approach of MPC.


\section{Future Work}
\label{future_work}
Even though we have shown relatively basic functionality of RDTs and validated their general performance, we have only been able to do so on simulated models.
An obvious next step would thus be to control a real-world, physical TT.
This could involve either building a physical incubator following the detailed instructions in Feng \textit{et al.}~\cite{feng_et_al_2022} or accessing theirs remotely.
Additionally, we are developing the necessary tooling to translate instance models and their trigger-action-rules from the popular Home Assistant platform into our own, which would subsequently allow us to experiment more easily with smart home models.

This highlights the largest obstacle to deploying the RDT: although obtaining a model of a smart home is a simple matter of inspection and conversion into an ontological instance model, deriving the necessary FMUs will require substantial expert-user interaction.
We envision that RDTs could integrate with Talasila \textit{et al.}'s ``Digital Twin as a Service platform''~\cite{talasila_et_al_2025}.
This platform already provides substantial features for model management, a user interface, and support for different stakeholders in DT's life-cycle.
Similar challenges hold for combined model and system maintenance as the TT evolves, forcing the DT to adapt.

The current implementation of the simulation tree generation only works for limited value domains of actuations that, together with \textit{:Effect} annotations and reactive heuristics serve as a first stop to prune the search space, but do not scale to large value domains.
For these, we will need a different strategy that dynamically explores a limited subset of values for such a parameter, e.g.\ through a branch-and-bound algorithm.
We would expect that such an algorithm dynamically generates direct child-simulations of the current state,
and uses feedback from the simulation of these children to decide whether to generate an alternative child, or proceed to the next prediction-level with (one or more of) those that it has already generated.
However, we expect that this feedback-mechanism will require us to some degree to rearchitecture the generation of the simulation tree.

It is clear that we have to replace the heavy round trip through Apache Jena's inference engine, which is where execution spends most of its time.
We plan to use in-memory instances of the Apache Jena Fuseki\cite{fuseki} server instead, which should drastically reduce execution time.

There are obvious improvement possibilities to the functionality of the concept.
As mentioned in Sections \ref{validation_m370_case-based_reasoning_results} and \ref{discussion}, there are currently discrepancies between when different kinds of properties are observed in a given cycle.
Since some properties (e.g., actuator-dependent ones) require the execution of certain phases of the loop before being updated, our concept also highlights a potential shortcoming of the MAPE-K architecture.
In general, because reading property values has essentially no impact on the TT, it may be useful to consider decoupling data observation into separate cycles from those of planning and executions, thus allowing for more frequent up-to-date information and a more interruptible and dynamic decision-making process.
This would also enable the implementation of a kind of decision lag mitigation \cite{walton_et_al_2024} mechanism to combat substantial changes in TT conditions that may occur during particularly long Analyze and Planning phases, in which an RDT's chosen actions are no longer pertinent to its current property values.
Furthermore, even though our current solution only allows for a single property value deviation tolerance for all instance model properties, we see a likelihood in domain experts requiring different tolerances for different properties, which may use entirely different value scales (e.g., temperature vs. humidity).
Lastly, as demonstrated in Section \ref{validation_m370_core_results}, having longer prediction horizons is not necessarily beneficial with respect to optimizations.
Although having variable-length prediction horizons might seem pragmatic, more research is needed in determining their applicability.

Our prototype can trivially control a system in a reactive manner once FMUs and adapters for sensors and actuators are in place, as we have shown with our examples.
However, the real payoff comes from being able to specify future goals that the RDT can target through its planning.
To specify these goals, users need an expressive language to express them as calculations on observed values and derived properties, such as the accumulated energy cost.
We have implemented the foundations for such a language through the building blocks of our R\&S algorithm, as well as the annotations to maximize or minimize such derived properties.
There is no user-facing syntax for how to write such expressions in the model yet.
Instead, as part of our evaluation, we have manually built the corresponding expressions.
The system currently supports primitive data types but still lacks constructs for time, meaning that specifications such as ``the car must be charged by 7am'' or ``the heater should run (at least) 4 hours within a 24-hour window'' can neither be conveniently specified nor planned for.
In the future, such expressions could make use of a kind of signal temporal logic (STL) \cite{hutchison_monitoring_2004}, such as cumulative-time signal temporal logic (CT-STL) \cite{chen_cumulative-time_2025}, and combine with existing tools to assist with pruning decision trees and picking optimal decisions.

Additionally, the RDT is currently not aware of long-running, uninterruptible actuations.
In case such operations are initiated, state tracking is required to restrict the set of available actions until the device has completed its operation.
While getting from expressions-as-strings to their realization is a relatively straightforward software engineering task, providing a suitable integrated development environment for the user to prototype their models with editor support and type-checking of complex optimal conditions, simulating them on historic values etc.\ is a separate issue.

As we have seen in our experiments, it is not always easy to understand why a particular set of decisions has been chosen, as this is encoded in intermediate calculations and in the R\&S algorithm.
We are currently working on a visualization component that allows users to explore the tree of possible decisions with their predicted intermediate states.
We anticipated that insight into the various components of the R\&S algorithm, as opposed to only the overall scores or decisions, will be necessary to increase transparency of the behavior of the RDT.

With the advent of generative AI, we could explore if the static structure of a smarthome with its sensors and actuators per room plus other simple meta-data provided by a non-expert user are sufficient to derive suitable skeleton FMUs for common civil ingeneering-scenarios such as heating, ventilation and optimal conditions.
Investigating the different smarthome platforms might also prove a reasonable starting point to evaluate the different ontologies against each other, in particular SAREF and SOSA/SSN.
So far, there are only limited comparison per domain, see e.g.\ Poveda{-}Villal{\'{o}}n \textit{et al.}~\cite{DBLP:conf/semweb/Poveda-Villalon18} for an analysis in the agricultural domain,
and Moreira \emph{et al.}~\cite{DBLP:conf/i-semantics/MoreiraDPSWSPGP17} attempt to answer this question by analysing mapping between ontologies.

The current implementation is relatively brittle with its interplay between the input models, the inference phase (through Apache Jena rules), and model queries with SPARQL in the MAPE-K loop.
Starting from consistency checks of instance models through SHACL~\cite{w3c_shacl_2017}, we could potentially go beyond well-formedness checks on models
and also consider checks on the consistency of our internal rules and queries.
\vsnote{Rephrased. I wanted to say that we can't evaluate if a query that the RDT uses internally actually makes sense, or if two queries together are fit for a particular purpose.}

We have not yet explored self-adaption: the RDT uses user-provided FMUs, and hence there is some flexibility in reconfiguring them as the system runs.
Although there exist different forms of drift (see e.g.\ Abbasi \textit{et al.}~\cite{DBLP:conf/models/AbbasiBPS24}), the easiest target would be data drift, e.g.\ due to the passing of the seasons, etc.
A separate component could detect an `anomaly' and provide reconfiguration actions to the involved FMUs.
In the case of the incubator and office room M370, this could be that the lid or a window is open through the deviation between the expected and observed temperatures.
However, designing anomaly detectors is out of the scope of this article.

\section{Conclusion}
\label{conclusion}
In this work, we present the concept of \textit{ruleless digital twins} (RDTs) as a means of DT decision-making enabled purely by declarative user specifications.
By combining standardized technologies, such as a semantic knowledge base (SSN), logical inference (Apache Jena rules), simulations (FMUs), and an autonomic computing architecture (MAPE-K), RDTs enact optimized sequences of actions under dynamically changing TT conditions, akin to the well-established \textit{ruleless} method of model predictive control (MPC).
As a result, RDTs relieve users of building and maintaining complex and conflict-prone rule-based decision-making models by only requiring declarative constraints and optimization specifications in the form of semantic instance model annotations.

Notably, our experiments against a virtual office room environment reveal that RDT core functionality outperforms a rule-based bang-bang controller through maintaining more stable room temperatures and saving more in terms of total energy costs incurred under dynamically changing spot pricing. 
Longer prediction horizons have shown to generally yield better optimizations, although this is not always the case due to their boundedness and inherent unpredictability of dynamic systems.
Because RDTs rely on exploring optimal decisions through constructing combinatorial decision trees, longer prediction horizons (deeper tree levels) do incur a performance hit, ultimately placing a prerequisite on users to determine their applicability to the problem domain at hand.

\bibliography{bibliography}

@article{kritzinger_et_al_2018,
	title = {Digital {Twin} in manufacturing: {A} categorical literature review and classification},
	volume = {51},
	copyright = {https://www.elsevier.com/tdm/userlicense/1.0/},
	issn = {24058963},
	shorttitle = {Digital {Twin} in manufacturing},
	doi = {10.1016/j.ifacol.2018.08.474},
	language = {en},
	number = {11},
	urldate = {2026-02-17},
	journal = {IFAC-PapersOnLine},
	author = {Kritzinger, Werner and Karner, Matthias and Traar, Georg and Henjes, Jan and Sihn, Wilfried},
	year = {2018},
	pages = {1016--1022}
}

@misc{digital_twin_consortium,
	title = {Definition of a {Digital} {Twin}},
	url = {https://www.digitaltwinconsortium.org/initiatives/the-definition-of-a-digital-twin/},
	language = {en-US},
	urldate = {2026-02-17},
	journal = {Digital Twin Consortium},
	author = {{Digital Twin Consortium}}
}

@incollection{larsen_et_al_2024,
	address = {Cham},
	title = {Engineering {Digital} {Twins} for {Cyber}-{Physical} {Systems}},
	isbn = {978-3-031-66718-3 978-3-031-66719-0},
	doi = {10.1007/978-3-031-66719-0\_1},
	language = {en},
	urldate = {2026-02-17},
	booktitle = {The {Engineering} of {Digital} {Twins}},
	publisher = {Springer International Publishing},
	author = {Larsen, Peter Gorm and Fitzgerald, John and Gomes, Cláudio},
	editor = {Fitzgerald, John and Gomes, Cláudio and Larsen, Peter Gorm},
	year = {2024},
	pages = {3--17}
}

@misc{azeau,
	title = {Femyou --- loading and running fmi-standard FMUs in dotnet core},
	copyright = {MIT},
	url = {https://codeberg.org/SELab_HVL/vsto-Femyou},
	urldate = {2026-09-27},
	author = {Azeau, Olivier and Stolz, Volker},
	_note = {MIT License, original-date: 2020-06-21T19:13:34Z}
}

@inproceedings{compagnucci_et_al_2023,
	address = {Västerås, Sweden},
	title = {Supporting {Digital} {Twins} {Systems} {Integrating} the {MERODE} {Approach}},
	copyright = {https://doi.org/10.15223/policy-029},
	isbn = {979-8-3503-2498-3},
	doi = {10.1109/MODELS-C59198.2023.00079},
	urldate = {2026-02-17},
	booktitle = {2023 {ACM}/{IEEE} {International} {Conference} on {Model} {Driven} {Engineering} {Languages} and {Systems} {Companion} ({MODELS}-{C})},
	publisher = {IEEE},
	author = {Compagnucci, Ivan and Snoeck, Monique and Asensio, Estefanía Serral},
	month = oct,
	year = {2023},
	pages = {449--458}
}

@misc{etsi,
	title = {{SAREF}: the {Smart} {Applications} {REFerence} ontology},
	url = {https://saref.etsi.org/core/v4.1.1/},
	urldate = {2026-02-17},
	author = {{European Telecommunications Standards Institute}}
}

@inproceedings{DBLP:conf/semweb/Poveda-Villalon18,
  author       = {Mar{\'{\i}}a Poveda{-}Villal{\'{o}}n and
                  Quang{-}Duy Nguyen and
                  Catherine Roussey and
                  Christophe de Vaulx and
                  Jean{-}Pierre Chanet},
  editor       = {Maxime Lefran{\c{c}}ois and
                  Ra{\'{u}}l Garc{\'{\i}}a{-}Castro and
                  Amelie Gyrard and
                  Kerry Taylor},
  title        = {Ontological Requirement Specification for Smart Irrigation Systems:
                  {A} {SOSA/SSN} and {SAREF} Comparison},
  booktitle    = {Proc.\ 9th Intl.\ Semantic Sensor Networks Workshop (SSN@ISWC)},
  series       = {{CEUR} Workshop Proceedings},
  pages        = {1--16},
  publisher    = {CEUR-WS.org},
  address = {Bonn},
  year         = {2018}
}

@inproceedings{DBLP:conf/i-semantics/MoreiraDPSWSPGP17,
  author       = {Jo{\~{a}}o L. R. Moreira and
                  Laura Daniele and
                  Lu{\'{\i}}s Ferreira Pires and
                  Marten van Sinderen and
                  Katarzyna Wasielewska and
                  Pawel Szmeja and
                  Wieslaw Pawlowski and
                  Maria Ganzha and
                  Marcin Paprzycki},
  editor       = {Anna Fensel and
                  Laura Daniele and
                  Lora Aroyo and
                  Victor de Boer and
                  S{\'{a}}ndor Dar{\'{a}}nyi and
                  Omar Elloumi and
                  Ra{\'{u}}l Garc{\'{\i}}a{-}Castro and
                  Laura Hollink and
                  Oana Inel and
                  Gerard Kuys and
                  Maria Maleshkova and
                  Munir Merdan and
                  Albert Mero{\~{n}}o{-}Pe{\~{n}}uela and
                  Thomas Moser and
                  Felix Leif Keppmann and
                  Efstratios Kontopoulos and
                  Lodewijk Petram and
                  Enrico Scarrone and
                  Ruben Verborgh},
  title        = {Towards IoT Platforms' Integration Semantic Translations between {W3C}
                  {SSN} and {ETSI} {SAREF}},
  booktitle    = {Joint Proc.\ of SEMANTiCS 2017 Workshops co-located with the
                  13th Intl.\ Conf.\ on Semantic Systems (SEMANTiCS 2017)},
  series       = {{CEUR} Workshop Proceedings},
  publisher    = {CEUR-WS.org},
  address = {Bonn},
  year         = {2017}
}

@article{jiang_et_al_2021,
	title = {A {Review} on {Soft} {Sensors} for {Monitoring}, {Control}, and {Optimization} of {Industrial} {Processes}},
	volume = {21},
	copyright = {https://ieeexplore.ieee.org/Xplorehelp/downloads/license-information/IEEE.html},
	issn = {1530-437X, 1558-1748, 2379-9153},
	doi = {10.1109/JSEN.2020.3033153},
	number = {11},
	urldate = {2026-02-17},
	journal = {IEEE Sensors Journal},
	author = {Jiang, Yuchen and Yin, Shen and Dong, Jingwei and Kaynak, Okyay},
	month = jun,
	year = {2021},
	pages = {12868--12881}
}

@incollection{kamburjan_et_al_2021,
	address = {Cham},
	title = {Programming and {Debugging} with {Semantically} {Lifted} {States}},
	volume = {12731},
	isbn = {978-3-030-77384-7 978-3-030-77385-4},
	doi = {10.1007/978-3-030-77385-4\_8},
	language = {en},
	urldate = {2026-02-17},
	booktitle = {The {Semantic} {Web}},
	publisher = {Springer International Publishing},
	author = {Kamburjan, Eduard and Klungre, Vidar Norstein and Schlatte, Rudolf and Johnsen, Einar Broch and Giese, Martin},
	editor = {Verborgh, Ruben and Hose, Katja and Paulheim, Heiko and Champin, Pierre-Antoine and Maleshkova, Maria and Corcho, Oscar and Ristoski, Petar and Alam, Mehwish},
	year = {2021},
	series = {Lecture Notes in Computer Science},
	pages = {126--142}
}

@article{lehner_et_al_2025,
	title = {Model-driven engineering for digital twins: a systematic mapping study},
	volume = {24},
	issn = {1619-1366, 1619-1374},
	shorttitle = {Model-driven engineering for digital twins},
	doi = {10.1007/s10270-025-01264-7},
	language = {en},
	number = {5},
	urldate = {2026-02-17},
	journal = {Software and Systems Modeling},
	author = {Lehner, Daniel and Zhang, Jingxi and Pfeiffer, Jérôme and Sint, Sabine and Splettstößer, Ann-Kathrin and Wimmer, Manuel and Wortmann, Andreas},
	month = oct,
	year = {2025},
	pages = {1339--1377}
}

@article{michael_et_al_2024,
	title = {Explaining {Cyberphysical} {System} {Behavior} {With} {Digital} {Twins}},
	volume = {41},
	copyright = {https://ieeexplore.ieee.org/Xplorehelp/downloads/license-information/IEEE.html},
	issn = {0740-7459, 1937-4194},
	doi = {10.1109/MS.2023.3319580},
	number = {1},
	urldate = {2026-02-17},
	journal = {IEEE Software},
	author = {Michael, Judith and Schwammberger, Maike and Wortmann, Andreas},
	month = jan,
	year = {2024},
	pages = {55--63}
}

@article{musen_and_protege_2015,
	title = {The {Protégé} project: a look back and a look forward},
	volume = {1},
	issn = {2372-3483},
	shorttitle = {The Protégé project},
	doi = {10.1145/2757001.2757003},
	language = {en},
	number = {4},
	urldate = {2026-02-17},
	journal = {AI Matters},
	author = {Musen, Mark A. and {The Protégé Team}},
	month = jun,
	year = {2015},
	pages = {4--12}
}

@incollection{paesler_et_al_2022,
	address = {Cham},
	title = {A {Formal} {Model} of {Metacontrol} in {Maude}},
	volume = {13701},
	isbn = {978-3-031-19848-9 978-3-031-19849-6},
	doi = {10.1007/978-3-031-19849-6\_32},
	language = {en},
	urldate = {2026-02-17},
	booktitle = {Leveraging {Applications} of {Formal} {Methods}, {Verification} and {Validation}. {Verification} {Principles}},
	publisher = {Springer International Publishing},
	author = {Päßler, Juliane and Aguado, Esther and Silva, Gustavo Rezende and Tarifa, Silvia Lizeth Tapia and Corbato, Carlos Hernández and Johnsen, Einar Broch},
	editor = {Margaria, Tiziana and Steffen, Bernhard},
	year = {2022},
	series = {Lecture Notes in Computer Science},
	pages = {575--596}
}

@article{talasila_et_al_2025,
	title = {Composable digital twins on {Digital} {Twin} as a {Service} platform},
	volume = {101},
	issn = {0037-5497, 1741-3133},
	doi = {10.1177/00375497241298653},
	language = {en},
	number = {3},
	urldate = {2026-02-17},
	journal = {SIMULATION},
	author = {Talasila, Prasad and Gomes, Cláudio and Vosteen, Lars B and Iven, Hannes and Leucker, Martin and Gil, Santiago and Mikkelsen, Peter H and Kamburjan, Eduard and Larsen, Peter G},
	month = mar,
	year = {2025},
	pages = {287--311}
}

@inproceedings{DBLP:conf/models/AbbasiBPS24,
  author       = {Faima Abbasi and
                  Pierre Brimont and
                  C{\'{e}}dric Pruski and
                  Jean{-}S{\'{e}}bastien Sottet},
  editor       = {Manuel Wimmer and
                  Alexander Egyed and
                  Beno{\^{\i}}t Combemale and
                  Marsha Chechik},
  title        = {Understanding Semantic Drift in Model Driven Digital Twins},
  booktitle    = {Proc.\ {ACM/IEEE} 27th Intl.\ Conf.\ on Model
                  Driven Engineering Languages and Systems, {MODELS} Companion 2024},
  pages        = {419--430},
  publisher    = {{ACM}},
  address      = {NY},
  year         = {2024},
  doi          = {10.1145/3652620.3688256},
}

@misc{apache_software_foundation,
	title = {Apache {Jena} - {Home}},
	url = {https://jena.apache.org/},
	urldate = {2026-02-17},
	author = {{The Apache Software Foundation}}
}

@misc{fuseki,
	title = {Apache {Jena} {Fuseki}},
	url = {https://jena.apache.org/documentation/fuseki2/},
	urldate = {2026-09-28},
	author = {{The Apache Software Foundation}}
}

@misc{modelica_association,
	title = {Functional {Mock}-up {Interface}},
	url = {https://fmi-standard.org/},
	urldate = {2026-02-17},
	author = {{The Modelica Association}}
}

@misc{w3c_sparql_2013,
	title = {{SPARQL} 1.1 {Query} {Language}},
	url = {https://www.w3.org/TR/sparql11-query/},
	urldate = {2026-02-17},
	author = {{The World Wide Web Consortium}},
	month = mar,
	year = {2013}
}

@misc{w3c_and_ogc_2017,
	title = {Semantic {Sensor} {Network} {Ontology}},
	url = {https://www.w3.org/TR/vocab-ssn/},
	urldate = {2026-02-17},
	author = {{The World Wide Web Consortium} and {Open Geospatial Consortium} and {Rob Atkinson} and {Raúl García-Castro} and {Joshua Lieberman} and {Claus Stadler}},
	month = oct,
	year = {2017}
}

@misc{vesse_et_al,
	title = {{dotNetRDF}},
	url = {https://dotnetrdf.github.io/},
	language = {en},
	urldate = {2026-02-17},
	author = {{Rob Vesse} and {Ron Michael Zettlemoyer} and {Khalil Ahmed} and {Graham Moore} and {Tomasz Pluskiewicz} and {Samu Lang}}
}

@article{galvez_del_postigo_gallego_et_al_2025,
	title = {Decision {Support} {Within} {Digital} {Twins} in {Manufacturing} {Ecosystems}: {A} {Review}},
	volume = {7},
	issn = {2516-8398, 2516-8398},
	shorttitle = {Decision {Support} {Within} {Digital} {Twins} in {Manufacturing} {Ecosystems}},
	doi = {10.1049/cim2.70041},
	language = {en},
	number = {1},
	urldate = {2026-02-17},
	journal = {IET Collaborative Intelligent Manufacturing},
	author = {Gálvez Del Postigo Gallego, Lucía and Lazarova‐Molnar, Sanja and Del Real Torres, Alejandro and Acevedo Galicia, Luis E.},
	month = jan,
	year = {2025},
	pages = {e70041}
}

@article{coelho_et_al_2021,
	title = {Simulation-based decision support tool for in-house logistics: the basis for a digital twin},
	volume = {153},
	issn = {03608352},
	shorttitle = {Simulation-based decision support tool for in-house logistics},
	doi = {10.1016/j.cie.2020.107094},
	language = {en},
	urldate = {2026-02-17},
	journal = {Computers \& Industrial Engineering},
	author = {Coelho, F. and Relvas, S. and Barbosa-Póvoa, A.P.},
	month = mar,
	year = {2021},
	pages = {107094}
}

@inproceedings{korth_et_al_2018,
	address = {Seattle, WA, USA},
	title = {Simulation-ready digital twin for realtime management of logistics systems},
	isbn = {978-1-5386-5035-6},
	doi = {10.1109/BigData.2018.8622160},
	urldate = {2026-02-17},
	booktitle = {2018 {IEEE} {International} {Conference} on {Big} {Data} ({Big} {Data})},
	publisher = {IEEE},
	author = {Korth, Benjamin and Schwede, Christian and Zajac, Markus},
	month = dec,
	year = {2018},
	pages = {4194--4201}
}

@incollection{listl_et_al_2021,
	address = {Cham},
	title = {Decision {Support} on the {Shop} {Floor} {Using} {Digital} {Twins}: {Architecture} and {Functional} {Components} for {Simulation}-{Based} {Assistance}},
	volume = 630,
	isbn = {978-3-030-85873-5 978-3-030-85874-2},
	shorttitle = {Decision {Support} on the {Shop} {Floor} {Using} {Digital} {Twins}},
	doi = {10.1007/978-3-030-85874-2\_30},
	language = {en},
	urldate = {2026-02-17},
	booktitle = {Advances in {Production} {Management} {Systems}. {Artificial} {Intelligence} for {Sustainable} and {Resilient} {Production} {Systems}},
	publisher = {Springer International Publishing},
	author = {Listl, Franz Georg and Fischer, Jan and Rosen, Roland and Sohr, Annelie and Wehrstedt, Jan C. and Weyrich, Michael},
	editor = {Dolgui, Alexandre and Bernard, Alain and Lemoine, David and Von Cieminski, Gregor and Romero, David},
	year = 2021,
	series = {IFIP Advances in Information and Communication Technology},
	pages = {284--292}
}

@article{walton_et_al_2024,
	title = {A unified digital twin approach incorporating virtual, physical, and prescriptive analytical components to support adaptive real-time decision-making},
	volume = {193},
	issn = {03608352},
	doi = {10.1016/j.cie.2024.110241},
	language = {en},
	urldate = {2026-02-17},
	journal = {Computers \& Industrial Engineering},
	author = {Walton, Ryan B. and Ciarallo, Frank W. and Champagne, Lance E.},
	month = jul,
	year = {2024},
	pages = {110241}
}

@inproceedings{yigal_et_al_2024,
	address = {Angers, France},
	title = {Automation of {Smart} {Homes} with {Multiple} {Rule} {Sources}},
	isbn = {978-989-758-699-6},
	shorttitle = {Automation of {Smart} {Homes} with {Multiple} {Rule} {Sources}},
	doi = {10.5220/0012556300003705},
	urldate = {2026-02-17},
	booktitle = {Proceedings of the 9th {International} {Conference} on {Internet} of {Things}, {Big} {Data} and {Security}},
	publisher = {SCITEPRESS - Science and Technology Publications},
	author = {Yigal, Hoffner and Eran, Kaufman and Amir, Avidan and Yovel, Elad and Harel, Fogel},
	year = {2024},
	pages = {40--52}
}

@misc{openhab,
	title = {Rules - {Introduction}},
	url = {https://www.openhab.org/docs/tutorial/rules_introduction.html#parts-of-a-rule},
	language = {en-US},
	urldate = {2026-02-17},
	author = {{openHAB}}
}

@incollection{soares_et_al_2021,
	address = {Cham},
	title = {Programming {IoT}-{Spaces}: {A} {User}-{Survey} on {Home} {Automation} {Rules}},
	volume = {12745},
	isbn = {978-3-030-77969-6 978-3-030-77970-2},
	shorttitle = {Programming {IoT}-{Spaces}},
	doi = {10.1007/978-3-030-77970-2\_39},
	language = {en},
	urldate = {2026-02-17},
	booktitle = {Computational {Science} – {ICCS} 2021},
	publisher = {Springer International Publishing},
	author = {Soares, Danny and Dias, João Pedro and Restivo, André and Ferreira, Hugo Sereno},
	editor = {Paszynski, Maciej and Kranzlmüller, Dieter and Krzhizhanovskaya, Valeria V. and Dongarra, Jack J. and Sloot, Peter M.A.},
	year = {2021},
	series = {Lecture Notes in Computer Science},
	pages = {512--525}
}

@article{manca_et_al_2019,
	title = {Supporting end-user debugging of trigger-action rules for {IoT} applications},
	volume = {123},
	issn = {10715819},
	doi = {10.1016/j.ijhcs.2018.11.005},
	language = {en},
	urldate = {2026-02-17},
	journal = {International Journal of Human-Computer Studies},
	author = {Manca, Marco and Paternò, Fabio and Santoro, Carmen and Corcella, Luca},
	month = mar,
	year = {2019},
	pages = {56--69}
}

@article{dalibor_et_al_2022,
	title = {A {Cross}-{Domain} {Systematic} {Mapping} {Study} on {Software} {Engineering} for {Digital} {Twins}},
	volume = {193},
	issn = {01641212},
	doi = {10.1016/j.jss.2022.111361},
	language = {en},
	urldate = {2026-02-17},
	journal = {Journal of Systems and Software},
	author = {Dalibor, Manuela and Jansen, Nico and Rumpe, Bernhard and Schmalzing, David and Wachtmeister, Louis and Wimmer, Manuel and Wortmann, Andreas},
	month = nov,
	year = {2022},
	pages = {111361}
}

@inproceedings{bolender_et_al_2021,
	address = {Madrid, Spain},
	title = {Self-{Adaptive} {Manufacturing} with {Digital} {Twins}},
	copyright = {https://ieeexplore.ieee.org/Xplorehelp/downloads/license-information/IEEE.html},
	isbn = {978-1-6654-0289-7},
	doi = {10.1109/SEAMS51251.2021.00029},
	urldate = {2026-02-17},
	booktitle = {2021 {International} {Symposium} on {Software} {Engineering} for {Adaptive} and {Self}-{Managing} {Systems} ({SEAMS})},
	publisher = {IEEE},
	author = {Bolender, Tim and Burvenich, Gereon and Dalibor, Manuela and Rumpe, Bernhard and Wortmann, Andreas},
	month = may,
	year = {2021},
	pages = {156--166} 
}

@article{magnanini_et_al_2021,
	title = {A {Digital} {Twin}-based approach for multi-objective optimization of short-term production planning},
	volume = {54},
	issn = {24058963},
	doi = {10.1016/j.ifacol.2021.08.077},
	language = {en},
	number = {1},
	urldate = {2026-02-19},
	journal = {IFAC-PapersOnLine},
	author = {Magnanini, Maria Chiara and Melnychuk, Oleksandr and Yemane, Anteneh and Strandberg, Hakan and Ricondo, Itziar and Borzi, Giovanni and Colledani, Marcello},
	year = {2021},
	pages = {140--145}
}

@inproceedings{newcomb_et_al_2017,
	address = {Vancouver BC Canada},
	title = {{IOTA}: a calculus for internet of things automation},
	isbn = {978-1-4503-5530-8},
	shorttitle = {{IOTA}},
	doi = {10.1145/3133850.3133860},
	language = {en},
	urldate = {2026-02-20},
	booktitle = {Proceedings of the 2017 {ACM} {SIGPLAN} {International} {Symposium} on {New} {Ideas}, {New} {Paradigms}, and {Reflections} on {Programming} and {Software}},
	publisher = {ACM},
	author = {Newcomb, Julie L. and Chandra, Satish and Jeannin, Jean-Baptiste and Schlesinger, Cole and Sridharan, Manu},
	month = oct,
	year = {2017},
	pages = {119--133}
}

@inproceedings{ur_et_al_2014,
	address = {Toronto Ontario Canada},
	title = {Practical trigger-action programming in the smart home},
	isbn = {978-1-4503-2473-1},
	doi = {10.1145/2556288.2557420},
	language = {en},
	urldate = {2026-02-25},
	booktitle = {Proceedings of the {SIGCHI} {Conference} on {Human} {Factors} in {Computing} {Systems}},
	publisher = {ACM},
	author = {Ur, Blase and McManus, Elyse and Pak Yong Ho, Melwyn and Littman, Michael L.},
	month = apr,
	year = {2014},
	pages = {803--812}
}

@misc{soares_et_al_scenarios_2021,
	title = {Smart {Home} {Automation} {Survey}},
	copyright = {Creative Commons Attribution 4.0 International, Open Access},
	doi = {10.5281/ZENODO.4531395},
	language = {en},
	urldate = {2026-02-25},
	publisher = {Zenodo},
	author = {Soares, Danny and Dias, João Pedro and Restivo, André and Ferreira, Hugo Sereno},
	collaborator = {Ferreira, Hugo Sereno and Restivo, André},
	month = feb,
	year = {2021}
}

@article{kabashkin_2025,
	title = {Ontology-{Driven} {Digital} {Twin} {Framework} for {Aviation} {Maintenance} and {Operations}},
	volume = {13},
	issn = {2227-7390},
	doi = {10.3390/math13172817},
	language = {en},
	number = {17},
	urldate = {2026-03-06},
	journal = {Mathematics},
	author = {Kabashkin, Igor},
	month = sep,
	year = {2025},
	pages = {2817}
}

@article{richards_2020,
	title = {Boyd’s {OODA} {Loop}},
	journal = {Necesse},
        volume = 5,
	issue = 1,
	pages = {142--165},
	url = {https://hdl.handle.net/11250/2683228},
	note = {\url{https://hdl.handle.net/11250/2683228}},
	author = {Richards, Chet},
	year = {2020}
}

@article{aamodt_and_plaza_1994,
	title = {Case-{Based} {Reasoning}: {Foundational} {Issues}, {Methodological} {Variations}, and {System} {Approaches}},
	volume = {7},
	issn = {09217126},
	shorttitle = {Case-{Based} {Reasoning}},
	doi = {10.3233/AIC-1994-7104},
	number = {1},
	urldate = {2026-03-08},
	journal = {AI Communications},
	author = {Aamodt, Agnar and Plaza, Enric},
	year = {1994},
	pages = {39--59}
}

@book{allemang_and_hendler_2011,
	address = {Waltham, MA},
	edition = {2nd},
	title = {Semantic {Web} for the working ontologist: effective modeling in {RDFS} and {OWL}},
	isbn = {978-0-12-385965-5},
	shorttitle = {Semantic {Web} for the working ontologist},
	doi = {10.1016/C2010-0-68657-3},
	publisher = {Morgan Kaufmann/Elsevier},
	author = {Allemang, Dean and Hendler, James A.},
	year = {2011}
}

@misc{w3c_manchester_owl_syntax,
	title = {{OWL} 2 {Web} {Ontology} {Language} {Manchester} {Syntax} ({Second} {Edition})},
	url = {https://www.w3.org/TR/owl2-manchester-syntax/},
	urldate = {2026-03-13},
	author = {{The World Wide Web Consortium}}
}

@misc{the_apache_software_foundation_inference,
	title = {Apache {Jena} - {Reasoners} and rule engines: {Jena} inference support},
	url = {https://jena.apache.org/documentation/inference/},
	urldate = {2026-03-19},
	author = {{The Apache Software Foundation}}
}

@article{kowalski_1974,
	address = {Amsterdam},
	title = {Predicate {Logic} as {Programming} {Language}},
	volume = {74},
	url = {https://www.doc.ic.ac.uk/~rak/papers/IFIP74.pdf},
	journal = {Information Processing},
	publisher = {North-Holland Publishing Company},
	author = {{Robert Kowalski}},
	year = {1974},
	pages = {569--574}
}

@article{genesereth_and_ginsberg_1985,
	title = {Logic programming},
	volume = {28},
	issn = {0001-0782, 1557-7317},
	doi = {10.1145/4284.4287},
	language = {en},
	number = {9},
	urldate = {2026-03-20},
	journal = {Communications of the ACM},
	author = {Genesereth, Michael R. and Ginsberg, Matthew L.},
	month = sep,
	year = {1985},
	pages = {933--941}
}

@misc{w3c_shacl_2017,
	title = {Shapes {Constraint} {Language} ({SHACL})},
	url = {https://www.w3.org/TR/shacl/},
	language = {en},
	urldate = {2026-03-25},
	author = {{The World Wide Web Consortium}},
	month = jul,
	year = {2017}
}

@inproceedings{blockwitz_et_al_2012,
	title = {Functional {Mockup} {Interface} 2.0: {The} {Standard} for {Tool} independent {Exchange} of {Simulation} {Models}},
	shorttitle = {Functional {Mockup} {Interface} 2.0},
	doi = {10.3384/ecp12076173},
	urldate = {2026-03-31},
	author = {Blockwitz, Torsten and Otter, Martin and Akesson, Johan and Arnold, Martin and Clauss, Christoph and Elmqvist, Hilding and Friedrich, Markus and Junghanns, Andreas and Mauss, Jakob and Neumerkel, Dietmar and Olsson, Hans and Viel, Antoine},
	month = nov,
	year = {2012},
	pages = {173--184}
}

@article{fritzson_et_al_2020,
	title = {The {OpenModelica} {Integrated} {Environment} for {Modeling}, {Simulation}, and {Model}-{Based} {Development}},
	volume = {41},
	issn = {0332-7353, 1890-1328},
	doi = {10.4173/mic.2020.4.1},
	language = {en},
	number = {4},
	urldate = {2026-04-02},
	journal = {Modeling, Identification and Control: A Norwegian Research Bulletin},
	author = {Fritzson, Peter and Pop, Adrian and Abdelhak, Karim and Ashgar, Adeel and Bachmann, Bernhard and Braun, Willi and Bouskela, Daniel and Braun, Robert and Buffoni, Lena and Casella, Francesco and Castro, Rodrigo and Franke, Rüdiger and Fritzson, Dag and Gebremedhin, Mahder and Heuermann, Andreas and Lie, Bernt and Mengist, Alachew and Mikelsons, Lars and Moudgalya, Kannan and Ochel, Lennart and Palanisamy, Arunkumar and Ruge, Vitalij and Schamai, Wladimir and Sjölund, Martin and Thiele, Bernhard and Tinnerholm, John and Östlund, Per},
	year = {2020},
	pages = {241--295}
}

@inproceedings{feng_et_al_2022,
	address = {San Diego, CA, USA},
	title = {Integration {Of} {The} {Mape}-{K} {Loop} {In} {Digital} {Twins}},
	copyright = {https://doi.org/10.15223/policy-029},
	isbn = {978-1-7138-5288-9},
	doi = {10.23919/ANNSIM55834.2022.9859489},
	urldate = {2026-04-08},
	booktitle = {2022 {Annual} {Modeling} and {Simulation} {Conference} ({ANNSIM})},
	publisher = {IEEE},
	author = {Feng, Hao and Gomes, Claudio and Gil, Santiago and Mikkelsen, Peter H. and Tola, Daniella and Larsen, Peter Gorm and Sandberg, Michael},
	month = jul,
	year = {2022},
	pages = {102--113}
}

@article{kephart_and_chess_2003,
	title = {The vision of autonomic computing},
	volume = {36},
	copyright = {https://ieeexplore.ieee.org/Xplorehelp/downloads/license-information/IEEE.html},
	issn = {0018-9162},
	doi = {10.1109/MC.2003.1160055},
	language = {en},
	number = {1},
	urldate = {2026-04-09},
	journal = {Computer},
	author = {Kephart, J.O. and Chess, D.M.},
	month = jan,
	year = {2003},
	pages = {41--50}
}

@misc{beckett_et_al_2014,
	title = {{RDF} 1.1 {Turtle}},
	url = {https://www.w3.org/TR/turtle/},
	urldate = {2026-04-10},
	author = {{David Beckett} and {Tim Berners-Lee} and {Eric Prud'hommeaux} and {Gavin Carothers}},
	month = feb,
	year = {2014}
}

@article{hong_et_al_2021,
	title = {Review on ranking and selection: {A} new perspective},
	volume = {8},
	issn = {2095-7513, 2096-0255},
	shorttitle = {Review on ranking and selection},
	doi = {10.1007/s42524-021-0152-6},
	language = {en},
	number = {3},
	urldate = {2026-02-18},
	journal = {Frontiers of Engineering Management},
	author = {Hong, L. Jeff and Fan, Weiwei and Luo, Jun},
	month = sep,
	year = {2021},
	pages = {321--343}
}

@book{camacho_model_2007,
	address = {London},
	edition = {Second Edition},
	series = {Advanced {Textbooks} in {Control} and {Signal} {Processing}},
	title = {Model {Predictive} control},
	isbn = {978-1-85233-694-3 978-0-85729-398-5},
	doi = {10.1007/978-0-85729-398-5},
	language = {eng},
	publisher = {Springer},
	author = {Camacho, E. F. and Bordons, Carlos},
	year = {2007}
}

@incollection{nikolaou_model_2001,
	title = {Model predictive controllers: {A} critical synthesis of theory and industrial needs},
	volume = 26,
	isbn = {978-0-12-008526-2},
	shorttitle = {Model predictive controllers},
	doi = {10.1016/S0065-2377(01)26003-7},
	language = {en},
	urldate = {2026-09-02},
	booktitle = {Advances in {Chemical} {Engineering}},
	publisher = {Elsevier},
	address = {USA},
	author = {Nikolaou, Michael},
	year = 2001,
	pages = {131--204}
}

@inproceedings{mcallister_advances_2022,
	address = {Atlanta, GA, USA},
	title = {Advances in {Mixed}-{Integer} {Model} {Predictive} {Control}},
	copyright = {https://doi.org/10.15223/policy-029},
	isbn = {978-1-6654-5196-3},
	doi = {10.23919/ACC53348.2022.9867869},
	urldate = {2026-09-02},
	booktitle = {2022 {American} {Control} {Conference} ({ACC})},
	publisher = {IEEE},
	author = {McAllister, Robert D. and Rawlings, James B.},
	month = jun,
	year = {2022},
	pages = {364--369}
}

@article{hariprasad_efficient_2017,
	title = {An {Efficient} and {Stabilizing} {Model} {Predictive} {Control} of {Switched} {Systems}},
	volume = {62},
	copyright = {https://ieeexplore.ieee.org/Xplorehelp/downloads/license-information/IEEE.html},
	issn = {0018-9286, 1558-2523},
	doi = {10.1109/TAC.2016.2613909},
	number = {7},
	urldate = {2026-09-06},
	journal = {IEEE Transactions on Automatic Control},
	author = {Hariprasad, K. and Bhartiya, Sharad},
	month = jul,
	year = {2017},
	pages = {3401--3407}
}

@inproceedings{spajic_and_stolz_2026,
	address = {Cham},
	title = {Ruleless {Digital} {Twins}},
	volume = {16421},
	isbn = {978-3-032-25551-8 978-3-032-25552-5},
	doi = {10.1007/978-3-032-25552-5_5},
	language = {en},
	urldate = {2026-09-07},
	booktitle = {From {Data} to {Models} and {Back}},
	publisher = {Springer Nature Switzerland},
	author = {Spajić, Ivan and Stolz, Volker},
	editor = {Lestingi, Livia and Salaün, Gwen and Requeno Jarabo, José Ignacio},
	year = {2026},
	series = {Lecture Notes in Computer Science},
	pages = {67--84}
}

@article{zheng_emergence_2022,
	title = {The emergence of cognitive digital twin: vision, challenges and opportunities},
	volume = {60},
	issn = {0020-7543, 1366-588X},
	shorttitle = {The emergence of cognitive digital twin},
	doi = {10.1080/00207543.2021.2014591},
	language = {en},
	number = {24},
	urldate = {2026-09-08},
	journal = {International Journal of Production Research},
	author = {Zheng, Xiaochen and Lu, Jinzhi and Kiritsis, Dimitris},
	month = dec,
	year = {2022},
	pages = {7610--7632}
}

@article{glimm_hermit_2014,
	title = {{HermiT}: {An} {OWL} 2 {Reasoner}},
	volume = {53},
	issn = {0168-7433, 1573-0670},
	shorttitle = {{HermiT}},
	doi = {10.1007/s10817-014-9305-1},
	language = {en},
	number = {3},
	urldate = {2026-09-12},
	journal = {Journal of Automated Reasoning},
	author = {Glimm, Birte and Horrocks, Ian and Motik, Boris and Stoilos, Giorgos and Wang, Zhe},
	month = oct,
	year = {2014},
	pages = {245--269}
}

@incollection{shull_selecting_2008,
	address = {London},
	title = {Selecting {Empirical} {Methods} for {Software} {Engineering} {Research}},
	isbn = {978-1-84800-043-8 978-1-84800-044-5},
	doi = {10.1007/978-1-84800-044-5_11},
	language = {en},
	booktitle = {Guide to {Advanced} {Empirical} {Software} {Engineering}},
	publisher = {Springer London},
	author = {Easterbrook, Steve and Singer, Janice and Storey, Margaret-Anne and Damian, Daniela},
	editor = {Shull, Forrest and Singer, Janice and Sjøberg, Dag I. K.},
	year = {2008},
	pages = {285--311}
}

@inproceedings{abburu_cognitwin_2020,
	address = {Cardiff, UK},
	title = {{COGNITWIN} – {Hybrid} and {Cognitive} {Digital} {Twins} for the {Process} {Industry}},
	copyright = {https://ieeexplore.ieee.org/Xplorehelp/downloads/license-information/IEEE.html},
	isbn = {978-1-7281-7037-4},
	doi = {10.1109/ICE/ITMC49519.2020.9198403},
	urldate = {2026-09-29},
	booktitle = {2020 {IEEE} {International} {Conference} on {Engineering}, {Technology} and {Innovation} ({ICE}/{ITMC})},
	publisher = {IEEE},
	author = {Abburu, Sailesh and Berre, Arne J. and Jacoby, Michael and Roman, Dumitru and Stojanovic, Ljiljana and Stojanovic, Nenad},
	month = jun,
	year = {2020},
	pages = {1--8}
}

@article{chen_cumulative-time_2025,
	title = {Cumulative-{Time} {Signal} {Temporal} {Logic}},
	volume = {24},
	issn = {1539-9087, 1558-3465},
	doi = {10.1145/3763237},
	language = {en},
	number = {5s},
	journal = {ACM Transactions on Embedded Computing Systems},
	author = {Chen, Hongkai and Zhang, Zeyu and Roy, Shouvik and Bartocci, Ezio and Smolka, Scott A. and Stoller, Scott and Lin, Shan},
	month = sep,
	year = {2025},
	pages = {1--23}
}

@incollection{hutchison_monitoring_2004,
	address = {Berlin, Heidelberg},
	title = {Monitoring {Temporal} {Properties} of {Continuous} {Signals}},
	volume = {3253},
	copyright = {http://www.springer.com/tdm},
	isbn = {978-3-540-23167-7 978-3-540-30206-3},
	doi = {10.1007/978-3-540-30206-3_12},
	language = {en},
	booktitle = {Formal {Techniques}, {Modelling} and {Analysis} of {Timed} and {Fault}-{Tolerant} {Systems}},
	publisher = {Springer Berlin Heidelberg},
	author = {Maler, Oded and Nickovic, Dejan},
	editor = {Hutchison, David and Kanade, Takeo and Kittler, Josef and Kleinberg, Jon M. and Mattern, Friedemann and Mitchell, John C. and Naor, Moni and Nierstrasz, Oscar and Pandu Rangan, C. and Steffen, Bernhard and Sudan, Madhu and Terzopoulos, Demetri and Tygar, Dough and Vardi, Moshe Y. and Weikum, Gerhard and Lakhnech, Yassine and Yovine, Sergio},
	year = {2004},
	note = {Series Title: Lecture Notes in Computer Science},
	pages = {152--166}
}

\end{document}